\documentclass[preprint,authoryear,12pt]{elsarticle}

\usepackage{epsfig}

\usepackage{amssymb}

\usepackage[ps2pdf,%
a4paper=true,%
breaklinks=true,%
colorlinks=true,%
pdfauthor={Bianchi et al.},%
pdftitle={The Ultraviolet Sky}%
]{hyperref}

\journal{Advances in Space Research}

\newcommand{\Teff}{\mbox{$T_{\rm eff}$}~}
\newcommand{\teff}{\mbox{$T_{\rm eff}$}}

\newcommand{\as}{\mbox{$^{\prime\prime}$~}}

\newcommand{\rv}{$R_V$~}

\newcommand{\aebv}{A$_{\lambda}$/$E_{(B-V)}$~}
\newcommand{\afebv}{A$_{FUV}$/$E_{(B-V)}$~}
\newcommand{\anebv}{A$_{NUV}$/$E_{(B-V)}$~}

\newcommand{\beqa}{\begin{eqnarray}}
\newcommand{\eeqa}{\end{eqnarray}}

\newcommand{\ebv}{\mbox{$E_{B\!-\!V}$}}
\newcommand{\eub}{\mbox{$E_{U\!-\!B}$}}

\begin{document}

\begin{frontmatter}



\title{The Ultraviolet Sky: An Overview  from the GALEX Surveys. 
\tnoteref{footnote1}}
\tnotetext[footnote1]{The catalogs presented in this paper are accessible on line.}


\author{Luciana Bianchi\corref{cor}\fnref{footnote2}}
\address{Dept. of Physics \& Astronomy, The Johns Hopkins University, 3400 N. Charles St.,  Baltimore, MD 21218, USA}
\ead{bianchi@pha.jhu.edu}

\ead[url]{http://dolomiti.pha.jhu.edu}

\author{Alberto Conti and Bernie Shiao} 
\address{Space Telescope Science Institute, 3700 S. Martin Drive, Baltimore, MD 21218, USA}


\begin{abstract}

The Galaxy Evolution Explorer (GALEX)  has performed the first surveys of the sky in the Ultraviolet (UV). Its legacy is an unprecedented database with 
more than 200 million source measurements  in far-UV (FUV) and near-UV (NUV), 
as well as wide-field imaging of extended objects, 
filling an important gap in our view of the sky across the electromagnetic spectrum.
The UV surveys  offer unique sensitivity for identifying and studying selected classes of astrophysical objects, both stellar and extra-galactic.  We examine the overall content and distribution of UV sources over the sky, and with magnitude and color. For this purpose, we have constructed final catalogs of UV sources with homogeneous quality, eliminating duplicate measurements of the same source. Such catalogs can facilitate a variety of  investigations on UV-selected samples,  as well as planning of observations with future missions.
We describe the criteria used to build the catalogs, their coverage and completeness.   We included observations in which  both the far-UV and near-UV detectors were exposed;  28,707 fields from the All-Sky Imaging survey (AIS) cover a unique area of 22,080 square degrees (after we restrict the catalogues to the central 1$^{\deg}$ diameter of the field), with a typical depth of $\sim$20/21~mag (FUV/NUV, in the AB mag system), and 3,008 fields from the Medium-depth Imaging Survey (MIS) cover a total of 2,251  square degrees at a depth of $\sim$22.7mag. 
The catalogs contain $\sim$71 and $\sim$16.6~million sources respectively. 
The density of hot stars reflects the Galactic structure,
and the number counts of both  Galactic and extra-galactic sources  are modulated by the Milky Way dust extinction, to which the UV data are very sensitive. 
\end{abstract}


\end{frontmatter}

\parindent=0.5 cm

\section{Introduction}
\label{s_intro}

For over eight years the Galaxy Evolution Explorer (GALEX, Martin et al. 2005, Morrissey et al. 2007,
Bianchi 2009, 2011) has surveyed the sky at ultraviolet (UV) wavelengths.
Other instruments have provided, and are providing, a variety of observations 
in this range of the electromagnetic spectrum, but no imaging  sky surveys had been performed  prior to GALEX. 
Not surprisingly,   many unexpected discoveries and results have emerged,
beyond the original science goals of the GALEX mission (e.g. Bianchi 2011 for
a partial review, Martin et al. 2007, Wyder et al. 2009, Thilker et al. 2007a, Bianchi 2012). 

A long-lasting heritage of GALEX  is  its database with over 200 million measurements 
of sources in two UV broad bands, far-UV (FUV, 1344-1786\AA) 
and near-UV (NUV, 1771-2831\AA),
as well as imaging covering most of the sky, and grism spectra for large subsamples. 
 This database is an 
unprecedented resource for stellar science, for studies of extragalactic objects such as star-forming galaxies and QSOs,
and for studies of interstellar medium and dust in particular, because UV fluxes are very sensitive 
to interstellar extinction by dust.
 It allows for the first time analysis  of UV-selected samples
of objects, or complementing with UV data measurements of large samples at longer wavelengths.

 On one hand, UV photometry combined with optical data critically increases the sensitivity to identify the
hottest stars, and to estimate their effective temperature, accounting for interstellar extinction. 
For example, the color difference between a \teff=50,000K and 20,000K star is
$\sim$1.5mag in FUV-{\it g}, but only  $\lesssim$0.4mag in U-B, and $\lesssim$0.15~mag in {\it g-r}, the latter being comparable to photometric
errors when all things are considered. The sensitivity gained by 
measurements at UV wavelengths is even more critical  
for discerning the extremely hot stars. 
In particular, hot  white dwarfs (WD) are elusive at all wavelengths except the UV, because of their very high \Teff and low 
optical luminosity. Bianchi et al. (2011a) extracted a first census  of hot WDs in the Milky Way from GALEX data and compared it with
Milky Way models  to test current assumptions on  late-stage evolution of intermediate-mass stars, independently from 
other methods.  Combined with optical measurements, the GALEX photometry also uniquely enables identification
of hot WDs in binaries (e.g. Sahai et al. 2008, Bianchi 2007 and 2013 in preparation).
As another example, QSOs with redshift $\approx$1 can be unambiguously
selected and separated from other classes of objects with combinations of UV and optical colors
(e.g. Hutchings et al. 2010). 

In this work we present an overview of the UV source distribution across the sky, and discuss 
the global content and characteristics of the final database obtained from the UV sky surveys.  
For this purpose we constructed catalogs of unique UV sources (removing duplicate measurements)
with homogeneous depth and quality. The catalogs are made available on line, and can support 
statistical studies of both Galactic and extragalactic objects. 

The paper is arranged as follows. 
In Section \ref{s_data} we describe the essential characteristics of the GALEX data; in   
Section \ref{s_catalogs} we explain the criteria used to construct the UV source catalogs and describe  
their content; in Section \ref{s_uvsky} we use the catalogs to present (for the first time) an overview of the 
UV sky, and in Section \ref{s_match} we summarize the major existing and upcoming optical surveys 
which provide useful corollary data  for the UV sources. Appendix A gives more details on the 
catalogs. 
Throughout this paper we refer to magnitudes in the ABmag system. 
Coefficients for transformation between AB and Vega magnitudes are given in Table 1
of Bianchi (2011) for the GALEX passbands and other major photometric systems.

\begin{figure}
\label{f_figure1}
\begin{center}
\vskip -3.25cm
\includegraphics*[width=10.cm]{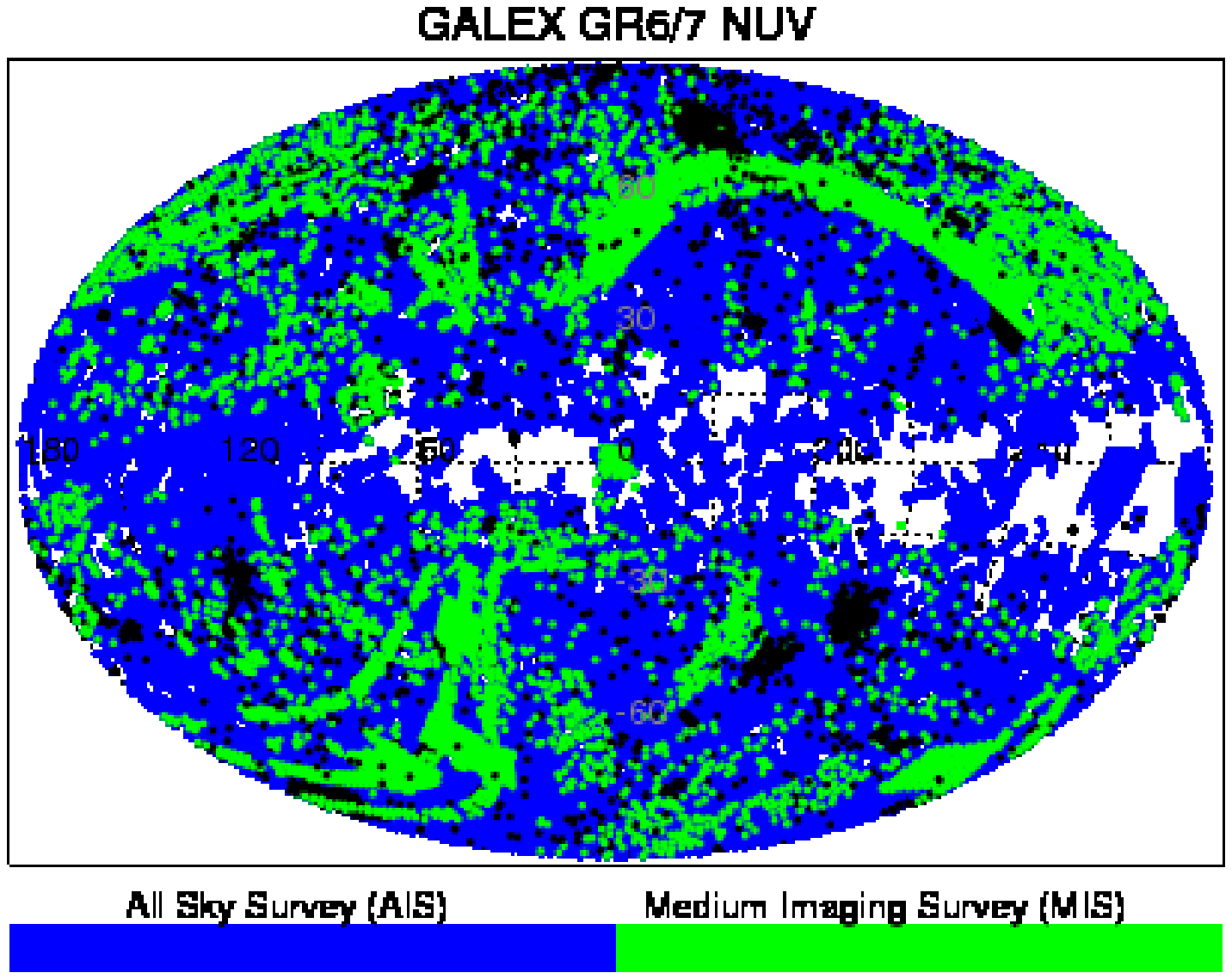}
\includegraphics*[width=10.cm]{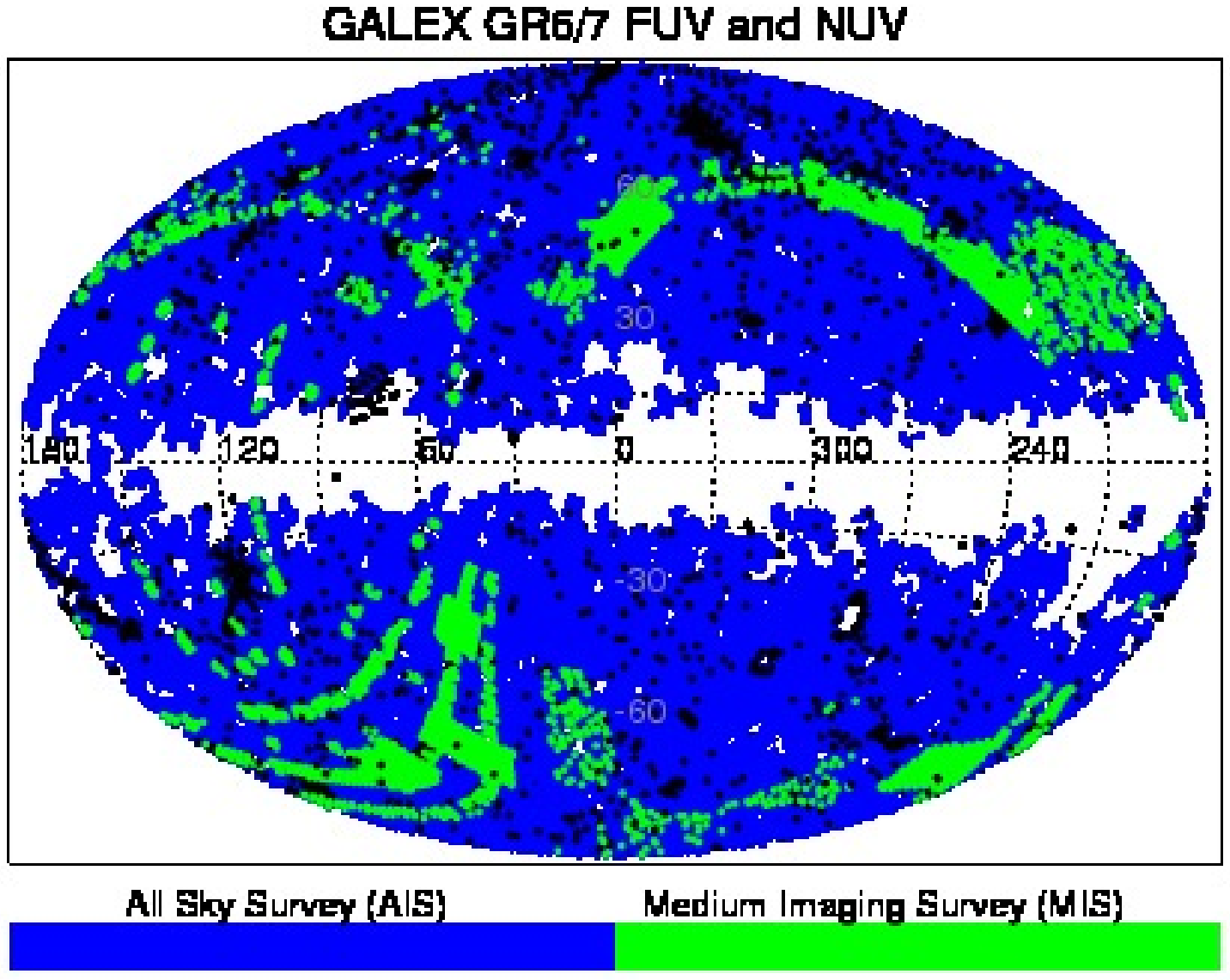}
\includegraphics*[width=5.cm]{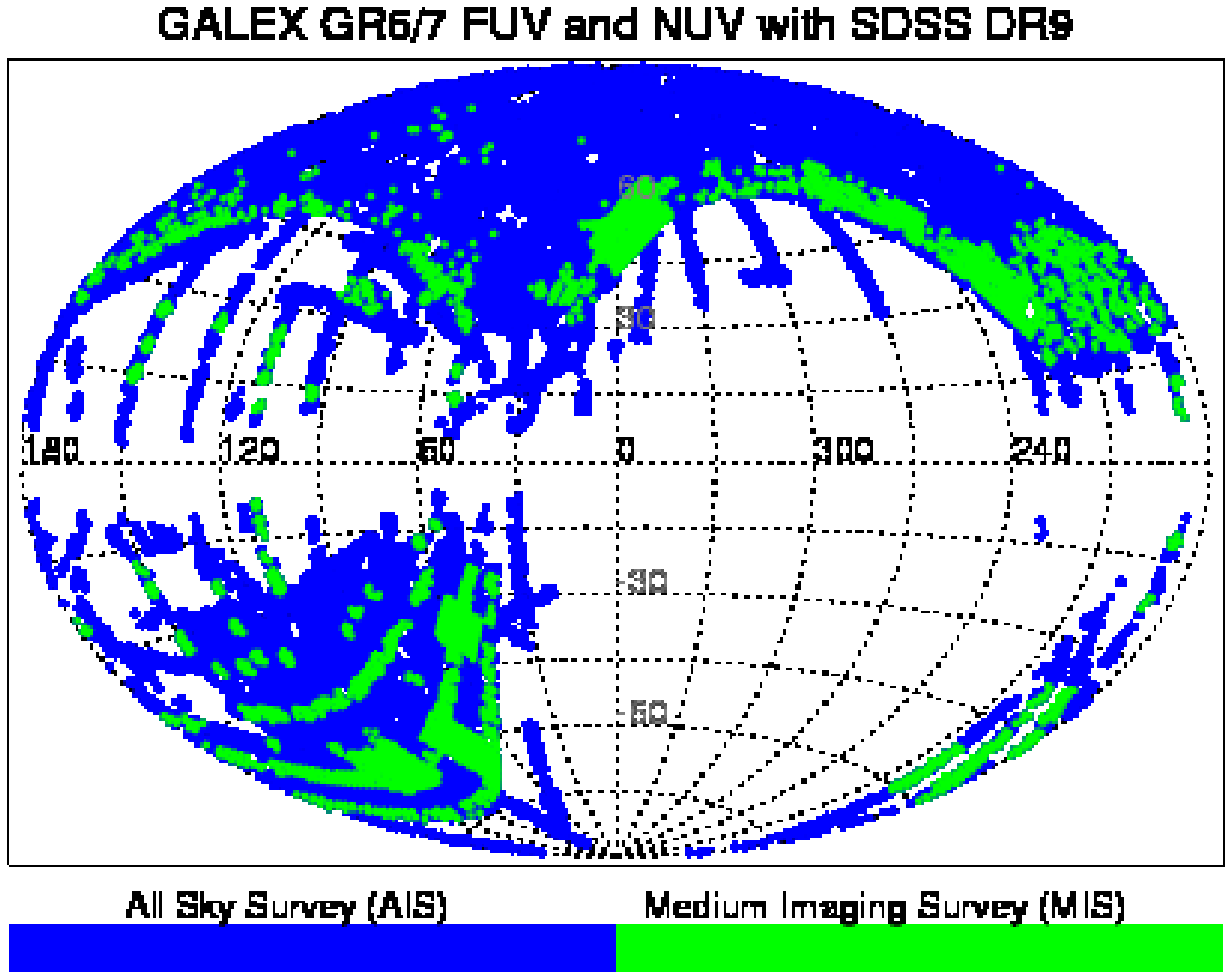}
\includegraphics*[width=5.cm]{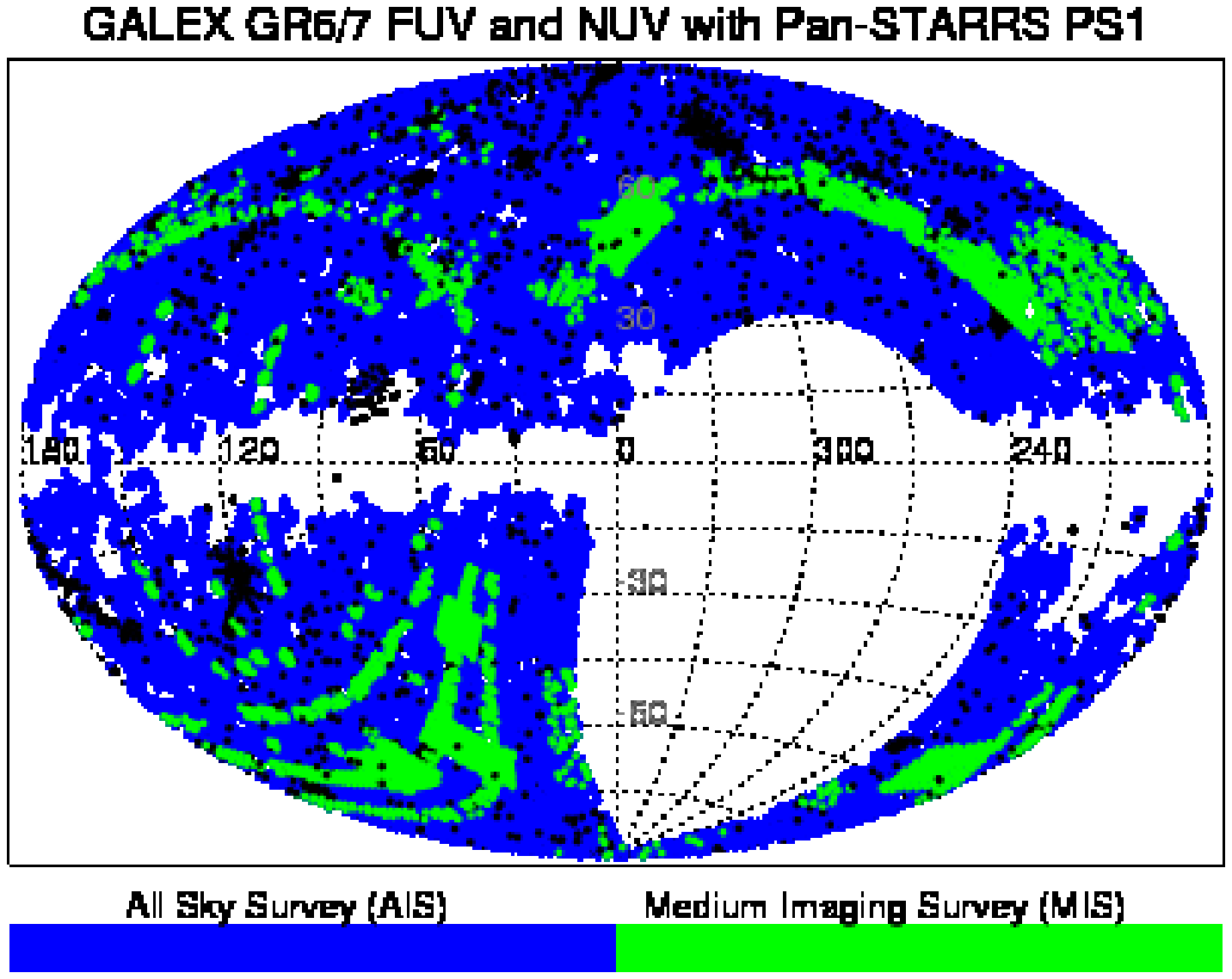}
\end{center}
\vskip -1.cm 
\caption{Sky coverage in Galactic coordinates of the final GALEX database, showing 
the  surveys AIS (blue) and  MIS (green). Observations from other surveys are in black. 
Top: fields with exposure in at least NUV; Middle: 
fields with exposure in both FUV and NUV detectors; Bottom: overlap between GALEX coverage (FUV \& NUV) and SDSS DR9 release (left), and
expected overlap 
with Pan-STARRS PS1 3$\pi$ survey soon to become available (right). }
\end{figure}

\section{The GALEX Data. The Sky surveys.} 
\label{s_data}

GALEX has performed sky surveys with different depth and coverage (see  Morrissey et al. 2007, Bianchi 2009)
in two Ultraviolet bands, FUV ($\lambda$$_{eff}$ $\sim$ 1528\AA, 1344-1786\AA) and  
 NUV ($\lambda$$_{eff}$ $\sim$ 2310\AA,  1771-2831\AA).
The two detectors  observed simoultaneously, thanks to a dichroic beam splitter, 
providing observations of the same field in the two bands at the same time.
The FUV detector stopped working in May 2009, and subsequent  GALEX observations only
provided NUV imaging.  

 The GALEX field of view is $\approx$1.2$^{\circ}$  diameter (1.28/1.24$^{\circ}$, FUV/NUV), and the spatial resolution is 
 $\approx$ 4.2/5.3\as (Morrissey e al. 2007); the FUV and NUV images for each observation are reconstructed 
from the photon list recorded by the photon-counting microchannel plate detectors, 
and are sampled with  virtual pixels of 1.5\as size. The standard pipeline processing then calculates a 
sky background image for each observation, and performs source photometry with a number of options.
The FUV and NUV detections are matched by the pipeline with a 3\as radius, to produce a merged list for 
each observation (Morrissey et al. 2007). 
All pipeline products are available from the MAST archive at http://galex.stsci.edu.  

 At the end  of 2012,  
the GALEX database contains 214,449,551 source measurements. Of these, 
210,691,504 are from observations with both FUV and NUV detectors on. 
Figure \ref{f_figure1} shows the sky coverage of all GALEX observations 
performed with both detectors on (middle map) and with the NUV detector regardless of FUV status (top map). 
After NASA ended the  GALEX mission in 2011, the observations resumed with support
from private institutions for about one year, and have now ended. 
The last part of the GALEX mission was devoted to extending the AIS survey towards 
the Galactic plane, largely inaccessible during the prime mission phase because of the 
many bright stars that could have damaged the detector. In addition, a survey of 
the Magellanic Clouds (MC), previously unfeasible for the same reason (brightness limits violation), has been completed 
after the main mission was concluded. These extensions include only
NUV measurements, as shown in Figure \ref{f_figure1}. Many of the recent MC observations are not yet in the public archive, they are presented in this book (Simons et al.).

 The number of sources detected in FUV is typically a small fraction 
(about 10\% or more)  of the total NUV detections,  
 hot stars and very blue galaxies being much rarer than 
``redder'' objects. 
The relative fraction varies with magnitude, and with Galactic latitude (see also Bianchi et al. 2011a,b). 
For any study of statistical samples or of classes 
of objects, it is necessary to distinguish whether a NUV source does not have an FUV measurement 
in the database because 
its FUV flux is below detection threshold, or because the FUV detector was not 
turned on when that field was observed. In either case the value of the FUV magnitude in the database is ``-999''.  
 Whether the FUV detector was on can be determined by 
checking that the exposure time in each detector was $>$0, which however 
requires relating each source in the database to the corresponding entry of its
original observation in the ``photoExtract'' table, using the tag {\it ``photoextractid''}, 
and retrieving the exposure times.  
Therefore, additional information is needed to separate actual FUV non-detections. 

 In addition, the database contains repeated measurements  for many sources: these 
are  useful for serendipitous variability searches, however individual sources must 
be counted only once for the purpose of statistical studies. 

 Finally, when the depth of a field or of an entire sample must be estimated, one must 
keep in mind that the FUV and NUV exposure times are identical in most observations, but differ in some cases
(see Table 1). 

To present an overview of the UV sky as afforded by GALEX at the end of its mission, and to  
facilitate studies of UV-selected samples or entire classes of objects, in which the FUV$-$NUV color is often 
involved (or a  UV$-$optical color), we have constructed ``clean'' catalogs of sources
from all GALEX observations when both detectors were exposed. 
We removed any duplicate observation of the same source, to obtain  ``unique source''
catalogs, and excluded sources on the outer rim of the fields since these contain many
artifacts, and photometry and astrometry are less accurate near the edge of the field.

\begin{table}[h]
\caption{The UV Surveys} 
\begin{tabular}{lrrrrrr}
\hline
Survey & \# fields          &  \# fields          & Tot. NUV exp. &  Tot. FUV exp.  & Tot. NUV exp. \\
       &  NUV on  &  FUV,NUV on   &   NUV on    &  FUV,NUV on & FUV,NUV on\\
\hline
Any &        43,717 &        33,858 & 39,113,488. & 21,205,228. & 29,441,882. &  \\ 
AIS &        34,207 &        28,707 &  6,470,830. &  4,769,754. &  5,717,310. &  \\ 
MIS &         6,489 &         3,008 & 11,790,231. &  5,295,070. &  5,999,276. &  \\ 
\hline
\end{tabular}
\label{table1}
\noindent{\small Note: The first column (``NUV on'') counts observations when at least the NUV detector was on;
 the second column 
(``FUV,NUV on'') includes observations with both detectors on: in most of these cases FUV and 
NUV  exposures are identical. 
The total exposure time is given, in seconds, in the last three columns.}
\end{table}

\begin{figure}
\label{f_histexp}
\begin{center}
\includegraphics*[width=13.cm]{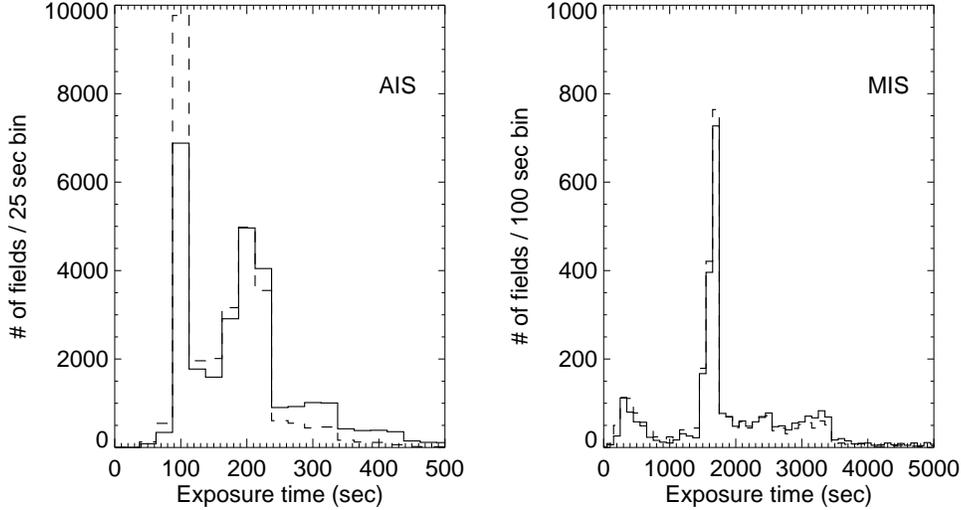}
\end{center}
\caption{Histograms of exposure times for the MIS and AIS surveys in the FUV detector (dashed line) and NUV detector (solid line).
Most AIS exposures are between 100 and 200 sec., and most MIS exposures are around the nominal $\sim$1,500~sec length.}
\end{figure}

\section{The Final Catalogs of unique UV sources}
\label{s_catalogs}


\begin{figure}
\label{f_maps}
\begin{center}
\includegraphics*[width=6.7cm]{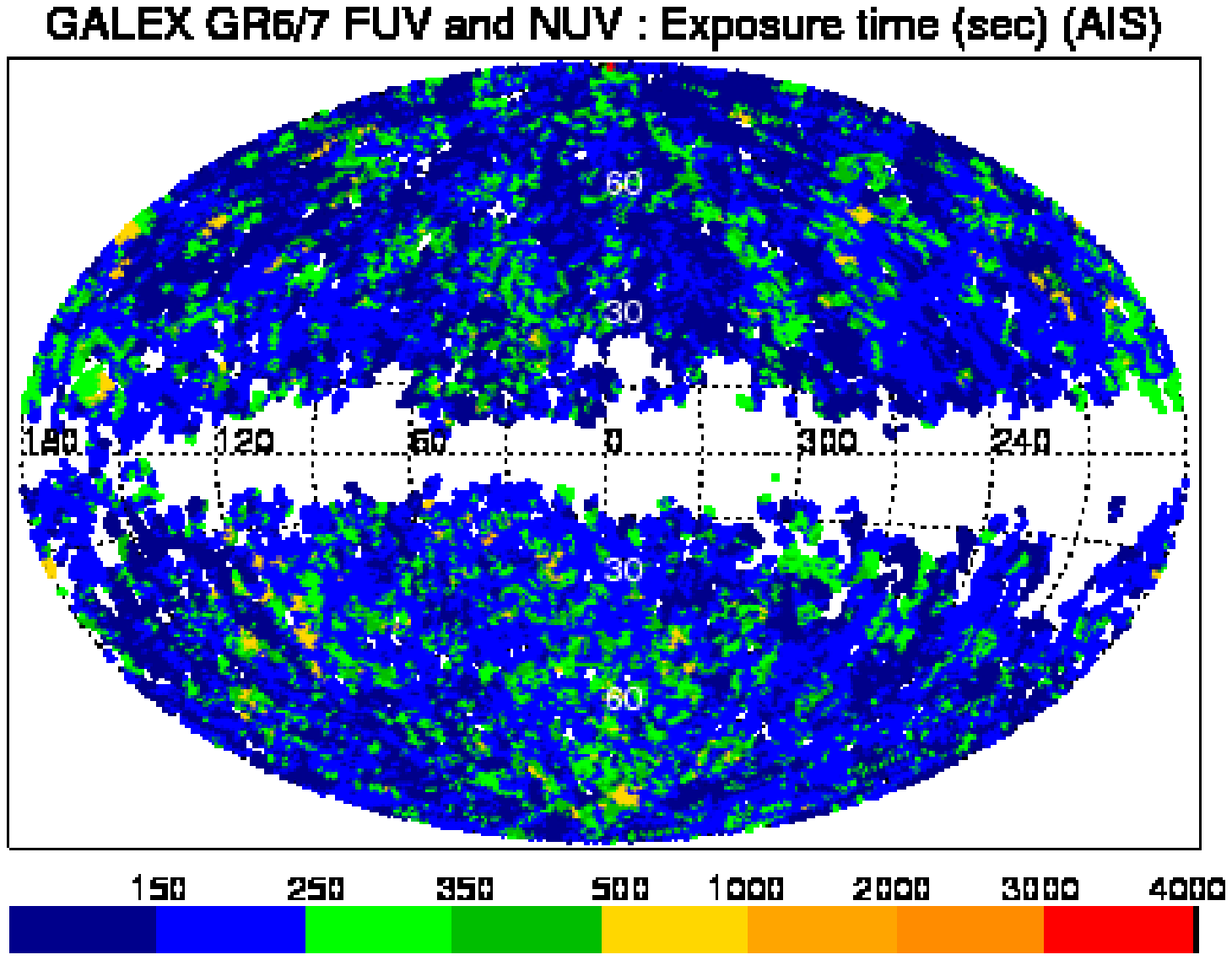}
\includegraphics*[width=6.7cm]{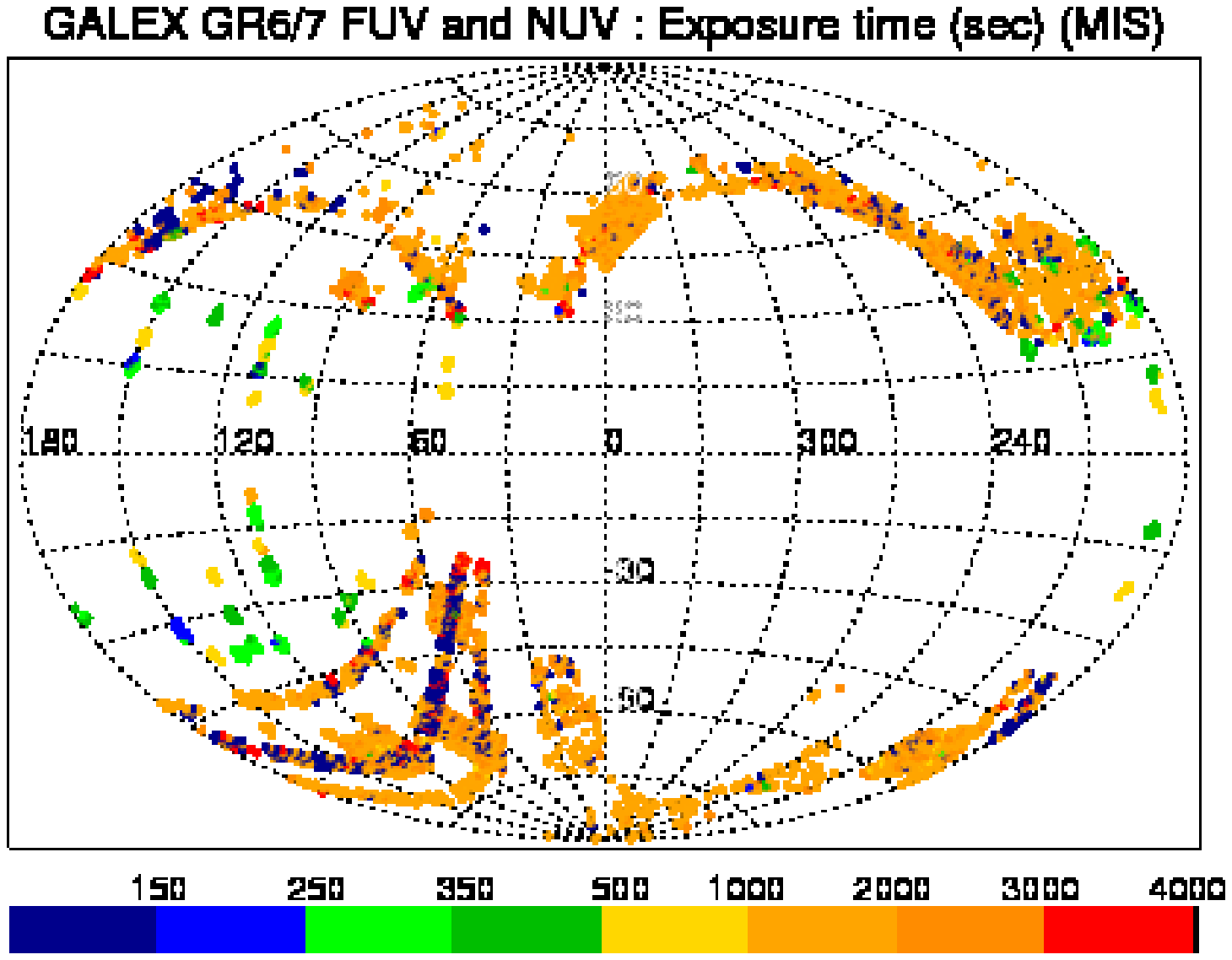}
\includegraphics*[width=6.7cm]{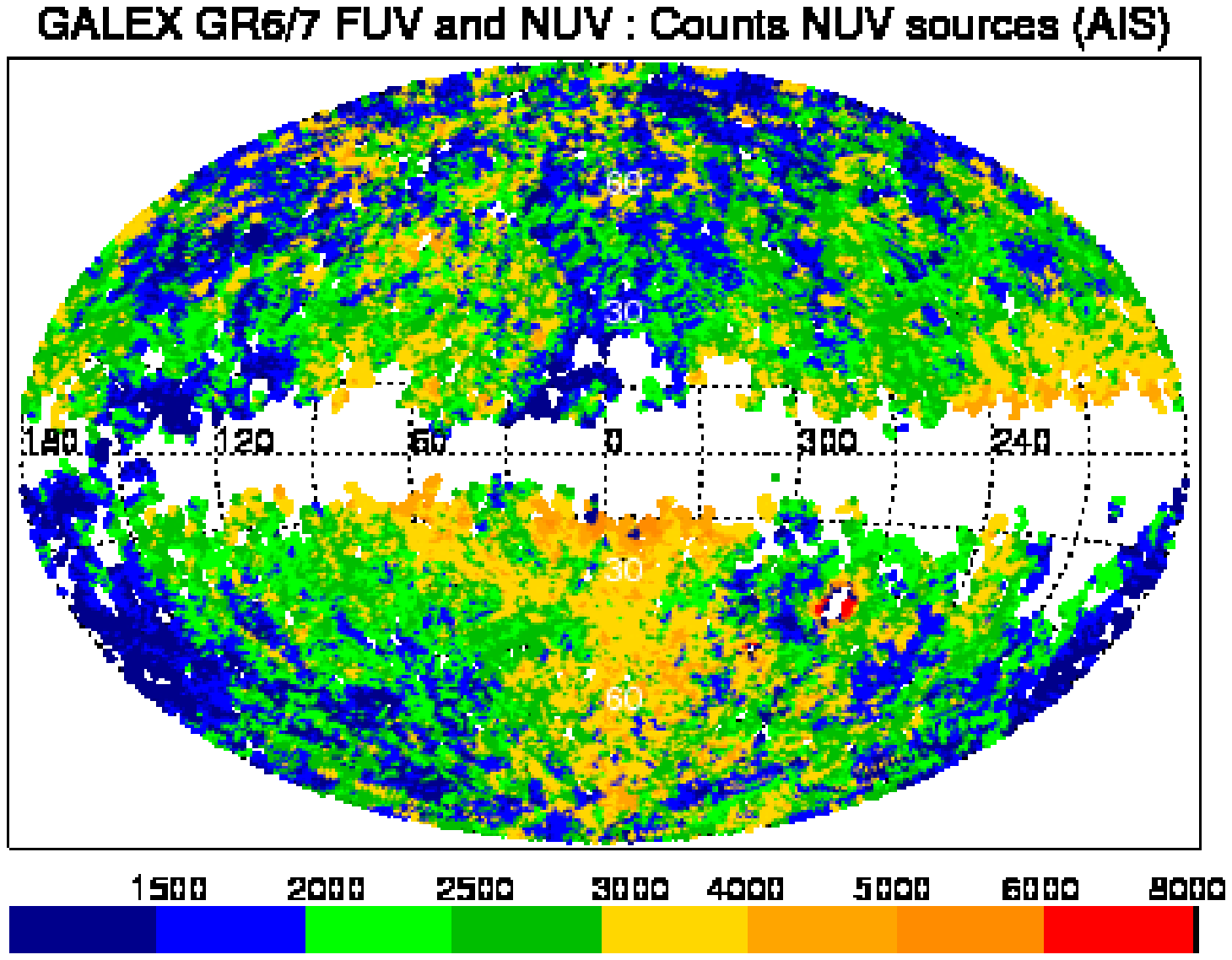} 
\includegraphics*[width=6.7cm]{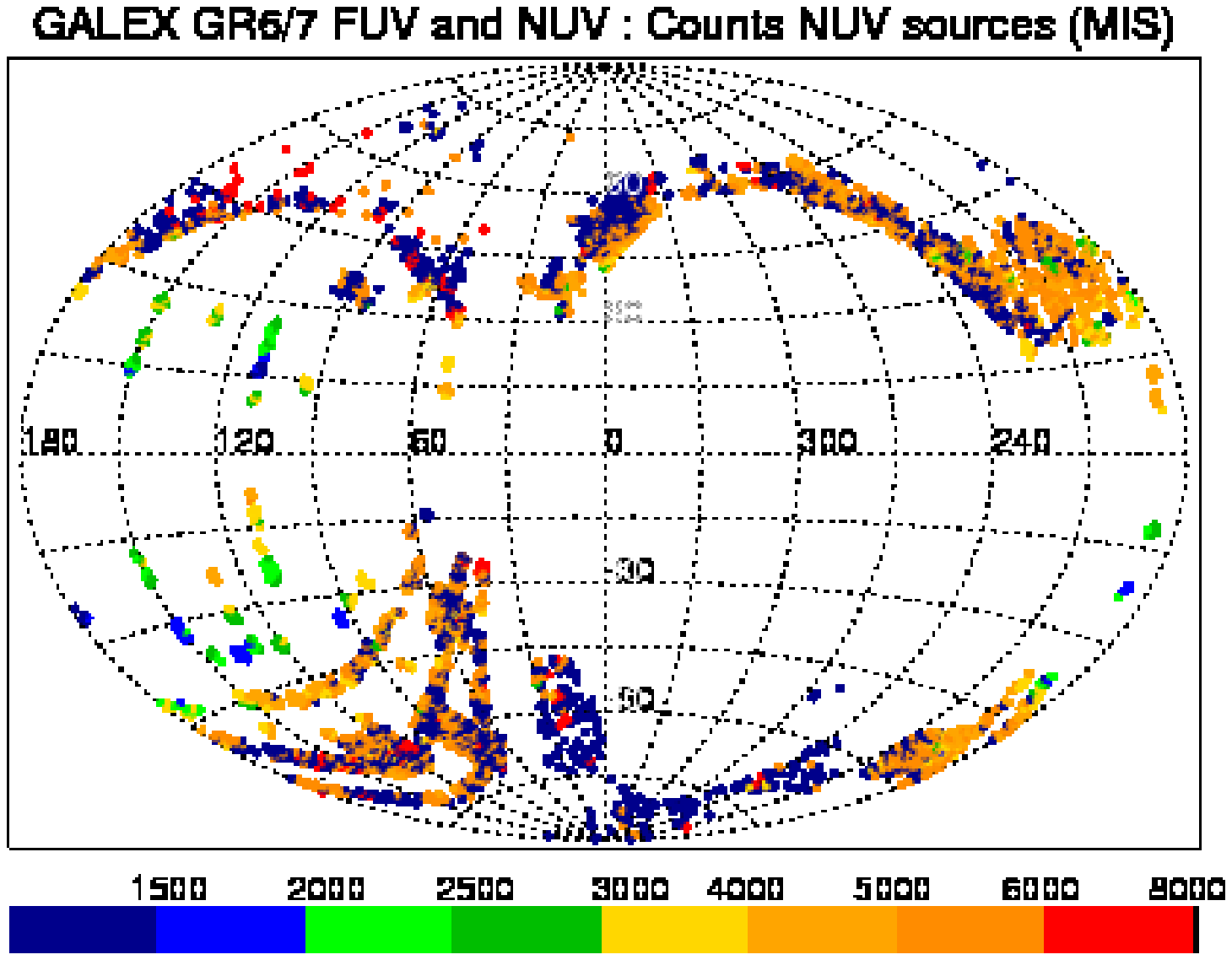} 
\end{center}
\vskip -1.cm 
\caption{
Maps in Galactic coordinates showing: (top) the distribution of NUV exposure time (in seconds);
(bottom)
the number of NUV detections (sources with NUV error$\leq$0.5, within 0.5degrees from each field center).
AIS is on the left, MIS on the right-side panels.  
The number of sources brighter than a given magnitude limit, in each location, depends on latitude for Galactic sources,
and on Milky Way extinction for both Galactic and extra-galactic sources. Extinction becomes significant
close to the Galactic plane  (see Bianchi et al. 2011a).  
Higher source counts are clearly seen in areas where exposure times are longer (compare top and bottom panel),
however conspicuous overdensity structures are seen also unrelated to longer exposures. 
}
\end{figure}

\subsection{Selection of Surveys. Sky Coverage}
\label{s_surveys}

For the purpose of an overall characterization of the UV sky, and to facilitate statistical studies, we constructed homogeneous catalogs from 
the  two GALEX surveys with the largest coverage,
AIS (All-Sky Imaging Survey, 
depth FUV/NUV $\sim$ 19.9/20.8~ABmag) 
and MIS (Medium-depth Imaging Survey (depth $\sim$22.7~mag in both FUV and NUV). 
 The magnitude limits given above correspond to the typical AIS and MIS exposures, which are  $\sim$150~sec and $\sim$1500~sec respectively, per survey design, 
but for some fields
the actual exposure times  are shorter or  longer, as shown in Figure 
\ref{f_histexp}.  
For AIS fields, the shortest exposure of an individual observation is 31~/~32~sec (FUV/NUV), 
and the longest is 708~/~1,133 sec (FUV /NUV); for MIS, the shortest exposure time is 
 39~/~55sec and the longest 15,541~/~22,220~sec for FUV/NUV.  These  cases are  rare. 
The distribution of exposure times is also shown at a glance across the sky in the top panels 
of Figure \ref{f_maps}.

 Observations in the GALEX database from surveys other than AIS and MIS  provide useful incremental coverage
for other purposes but
were not included in the catalogs presented here, in the interest of building two general catalogs of rather homogeneous 
quality. 
The ``Nearby Galaxies Survey'' (Gil de Paz et al. 2007, Bianchi et al. 2003) covered 436 fields with 
 exposure times comparable to MIS.
However, we did not include them because they specifically targeted extended galaxies, 
and most pipeline-defined sources are UV-bright regions in these galaxies (for which pipeline measurements are useless). 
These images are useful for studies of star formation across entire galaxies,
but require custom photometry (e.g., Kang et al. 2009, 
Efremova et al. 2011, Thilker et al. 2007b, Marino et al. 2010, 2011).
Observations from GI programs (1,322 fields) were not included because they have a wide range of exposures, but 
especially because they often targeted special objects such as star clusters, too crowded for the
pipeline to reliably extract individual sources (e.g. de Martino et al. 2008). 

 The alternative choice of including in our catalogs all observations divided in ranges of  
exposure time (similar to AIS and MIS exposure ranges), in principle preferable,  was
therefore discarded because of the many special  
objects, and consequently  unreliable photometric measurements, that would have been otherwise included.
Our restriction to designated ``AIS'' and ``MIS'' fields minimizes such cases, 
but does not entirely exclude them. 
 Although the largest and nearest galaxies were targeted in the Nearby Galaxies Survey,
 several extended galaxies (extended enough  that the pipeline may ``shred'' them and ingest  
measurements of individual emission peaks in the database)
are within the AIS and MIS footprints. When 
samples in wide areas are selected, these galaxies
must be weeded out. The exact criteria are better adjusted to the specific science goals, therefore
we did not apply any  cut to the catalogs that would have introduced ``holes'' in the coverage.
For most science applications of our catalogs, one can extract 
from the hyperLEDA database (http://leda.univ-lyon1.fr/)  a list of extended galaxies, 
 above certain dimensions  and brightness level, and flag UV sources within the
diameter of such galaxies in our catalogs, since 
samples may be contaminated by ``shredded-galaxy'' sources (e.g. Bianchi et al. 2007).  
If a given source of interest lies along the line of sight of an extended galaxy, the reliability of  its 
measurements can be checked from the catalog by looking at the many tags included, such as ellipticity
and size of the source,  by comparing PSF-, aperture-, and isophotal-photometry (all included in our catalog, 
see appendix A), and ultimately by examining the source in the original image data.

 Another important caveat concerns fields in or near the Magellanic Clouds, where the quality of 
 pipeline source detection and photometry 
is progressively degraded towards the more crowded  central fields. Carefully tuned and manually checked photometry is needed
even in the periphery of these galaxies, and pipeline measurements are not reliable.  
A separate catalog with custom-photometry 
for these regions is in preparation (Thilker et al.); a partial version is presented in this book (Simons et al. 2013).
For example, the pipeline may interpret two nearby sources as one extended source (see Simons et al. 2013 for an illustration of this problem).
Again, we did not exclude these areas from the present catalogs constructed from pipeline measurements, 
 to avoid introducing unnecessary (for some  purposes) boundaries and holes.
 
The same crowding issues apply to fields including Galactic star clusters. These should also be used with caution,
or better yet, excluded from statistical analyses of sources, since our catalogs are constructed from pipeline photometry for 
the purpose of homogeneity. 
These fields (when designated as MIS or AIS observations)
are included in the present catalogs for completeness, but again we stress
that for specific studies of crowded fields the pipeline measurements should be inspected carefully, and in many cases custom photometry is needed.

 For science goals not requiring statistical samples, such as searching for specific objects (for example the counterpart of an 
X-ray source), one may want to inspect the whole archive, in case the objects fall in 
portions of GALEX fields near the rim, that have been excluded from our catalogs
but may also be of some use if no better data are available. 
In such cases a direct search of the whole GALEX database is warranted, that will also turn up 
any existing repeated measurement for that source. 

While our catalogs of ``unique sources'' facilitate all studies involving source counts, 
the repeated measurements available in the GALEX database for a number of  sources 
 can be used for serendipitous variability searches.  A  catalog
of GALEX variables is presented by Conti et al. (2013), and specific searches for variables
in restricted areas or with restricted criteria have been performed, see
for example  Welsh et al. (2006, 2007, 2011),
 Wheatley et al. (2012), and 
Gezari et al. (2013,  
for GALEX sources matched with Pan-STARRS variables).

\subsection{Criteria for Constructing the Catalogs}
\label{s_criteria}

 The catalogs were constructed with the  criteria given below, following the method of 
 Bianchi et al. (2011a), where more details can be found, and of which the present catalogs
represent the updated, expanded,  and final version. 
\noindent
{\bf We retained in the catalogs only sources:}
\begin{itemize}
\item{{\bf from observations with both FUV and NUV detectors on.} This is useful for science analyses in which
 the fraction of sources with a given FUV-NUV color is of interest, or to estimate the fraction of sources with significant detection in both FUV and NUV over the total number of sources with NUV-only detection. More observations, 
taken with one of the two detectors turned  off (mostly FUV), can be found in the MAST database. 
Including observations where one detector was not exposed would bias any statistical analysis,
since the FUV magnitude of a NUV-detected source 
may appear as a non-detection (FUV=-999) either because the FUV detector was off, 
or the FUV detector was on but the FUV flux of that source was actually below detection threshold.}

\item{{\bf within the central 0.5 degrees radius of the field-of-view} ({\it fov\_radius} $\leq$ 0.5deg.), to avoid sources with poor photometry/astrometry near the edge, and rim artifacts. This restriction makes the catalogs useful for  analysis of source samples  with homogeneous quality, without great loss of area coverage (considering that overlap exists between several fields). Users interested in a particular source that happens to fall on the outermost edge of a GALEX field should obtain the measurements from the main database and carefully examine the quality.}
\item{{\bf with NUV magnitude errors $\leq$ 0.5mag}; this means all NUV detections are retained, regardless of detection in the FUV filter; see
Figures \ref{f_maps}, \ref{f_maps2}, 
 and \ref{f_histfn},  and Table 2 for coverage.  
The effect of error cuts on the resulting samples can be seen also from 
Figure 4 of Bianchi et al. 2011a, and Figures 2$-$4 of Bianchi et al. 2011b. }

\end{itemize}

\noindent
{\bf Removing Duplicate Measurements}. The general GALEX database
in the MAST archive  contains all existing measurements. For sources with repeated observations (e.g. where different fields overlap, or the same field was repeated), we removed duplicate measurements as follows, to produce a unique source catalog. GALEX sources within 2.5\as of each other, but from different observations, were considered duplicates. In such cases the object from the observation with the longest exposure time was retained, and - in cases of equal exposure times - the object closest to the center of the field  in its parent image. 

 The GALEX pipeline source detections uses a 3\as radius to match FUV sources to the NUV sources of the same observation. Our choice of a slightly smaller 
radius is based on initial tests for the early version of these catalogs (Bianchi et al. 2011a), as a good compromise between not excluding real sources 
and not retaining duplicate measurements. GALEX astrometry is usually more accurate than 2.5\as but the deblending of sources closer than 
this separation is not always robust due to the resolution. 
 
\subsection{Area Coverage of the Catalogs}
\label{s_area}

For any study involving density of sources in the sky (number per unit area), the exact area coverage of the 
catalog must be known. 
As we removed duplicate measurements of the same source, we  must also calculate the  area 
covered by the surveys accounting for 
overlaps, as well as 
for possible gaps, between fields. These may occur because of the tiling strategy, 
or because the actual pointing of an observation may be slightly off from the planned position, 
and also because we limited our catalogs to sources within the central 1$^{\circ}$ diameter of the GALEX field. 

We calculated the total area with the method 
of  Bianchi et al. (2011a): we divided the sky in small tesserae, 
 and added the 
areas of all tesserae which fall within 0.5$^{\circ}$ of every field center, checking that each tessera
is counted only once.  We obtain a total area  of 22,080 square degrees for the AIS catalog, and of 
2,253  square degree for the MIS catalog. Since we have restricted the catalogs to the inner 0.5$^{\circ}$ radius 
of each field, these areas imply that the remaining overlap among MIS fields is about 5\%, and 2\% for AIS. 
Had we not reduced the radius to 0.5$^{\circ}$, a larger area (by about 20\%  for MIS) would be gained, but many
artifacts and bad measurements would have  been included.   

 We found 3 fields in the MIS survey where both FUV and NUV detectors were exposed, but some
  issue caused NUV and FUV detections to not match, i.e. there are a number of 
sources with NUV measurements, all showing no FUV detection (FUV=-999), and viceversa
all FUV sources have NUV=-999. 
These are MIS  fields with {\it photoextractid} = 2462169510929498112, 6401235888134684672
and 3719131203803021312. When we exclude them, the total MIS area becomes 
2,251 
~square degrees.
The same problem was found for AIS fields with {\it photoextractid} = 6379923033125027840, 6381259965176217600, 6379711852804308992, 
6372041728408420352 and 6379571150749433856. 

 Because both gaps and overlaps between fields  exist,  the actual area coverage 
must be computed for each region of the sky where one desires to extract a sample, 
if the density of sources has to be estimated. 
Tables will be posted on the author's web site http://dolomiti.pha.jhu.edu/uvsky to 
facilitate  area calculations of custom-chosen regions.

\subsection{Content and Structure of the Catalog }
\label{s_content}

\ref{t_table2} gives 
the number of sources included in each catalog, both total and with various selections. 

An overall view of the density of UV sources and their characteristics across the sky is shown in Figures \ref{f_maps} (bottom) and \ref{f_maps2}.
 Figure \ref{f_histfn}  shows the distribution of  magnitude and color of the UV  sources. 
These are discussed in the next section. 

The catalogs (one for AIS, one for  MIS) contain several tags for every source, including position
(R.A., Dec., Galactic l,b), photometry measurements in FUV and NUV and their errors (``nuv\_mag'' and ``fuv\_mag'' are the
``best'' measurements as chosen by the pipeline, and preferable in most cases; other measurements are also
included, such as  PSF photometry, aperture photometry with different apertures, 
and Kron-like elliptical aperture magnitudes),
and parameters useful to retrieve the original image from which the photometry was extracted (tag {\it photoextractid}),
as well as artifact flags and extraction flags that can be used to eliminate spurious sources. 
The complete list of tags and their description is given in Appendix A. 

The catalogs can be downloaded from the author's web site: \\
http://dolomiti.pha.jhu.edu/uvsky , and will be also  available from the MAST web site,
and from the SIMBAD (Vizier)  database,
which allows VO-type queries including  cross-correlation with other catalogs in the
same database. 
An early version of similar catalogs (with less data coverage and fewer parameters) is  
VO-accessible with Vizier 
at: \\
http://vizier.u-strasbg.fr/viz-bin/VizieR-3?-source=II/312 , and from MAST at:\\ 
http://archive.stsci.edu/prepds/bianchi\_gr5xdr7/ .

\begin{table}[h]
\caption{Catalogs of unique UV sources$^a$}
\begin{tabular}{lrrrrrr}
\hline
Survey & {\tiny \# sources with error   $\leq$0.5mag}  & {\tiny \# sources  with error   $\leq$0.3mag }  & {\tiny \#  with error   $\leq$0.3mag}  \\
       &  {\tiny in NUV  /  in FUV \& NUV }         &  {\tiny in NUV  /  in FUV \& NUV }        &   {\tiny FUV-NUV$\leq$-0.1 / $\leq$0 } \\   
\hline
AIS    &   70,926,176 /  7,016,318 & 22,518,082 / 2,450,425 & 356,999 / 482,043  \\
MIS    &   16,613,047 /  3,986,646 &  6,552,960 / 2,053,537 & 135,320 / 202,875  \\
\hline
\end{tabular}
\label{t_table2}
\noindent{\small Note: The first number in columns 1 and 2 (``in NUV'') is the total number of sources detected in NUV,
regardless of FUV detection; the second number in columns 1 and 2, and the numbers in column 3, are source counts 
when the error cut is applied to both FUV and NUV measurements. The last columns give source counts also with a FUV-NUV cut. }
\end{table}

\subsection{Bright Object Limits}

We recall here that UV sources with a high countrate cause non-linearity to set in, 
beginning with a 10\% rolloff  at 109~counts~s$^{-1}$ for FUV and 311~counts~s$^{-1}$ for NUV.
These countrates correspond to FUV=13.73~ABmag ($\sim$1.53~e-13 erg~s$^{-1}$~cm$^{-2}$~\AA$^{-1}$)
and NUV=13.85 ($\sim$6.41~e-14 erg~s$^{-1}$~cm$^{-2}$~\AA$^{-1}$). A correction for non-linearity 
is applicable over a limited range, as shown by Morrissey et al.(2007; their figure 8), after which the measured 
countrates saturate and the true source flux is no longer recoverable. 
The bright-object limit during the major part of the mission was 30,000~counts~s$^{-1}$ per source,
corresponding to $\sim$9$^{th}$mag for NUV ($\sim$7~e-12 erg~s$^{-1}$~cm$^{-2}$~\AA$^{-1}$)
and 5,000~counts~s$^{-1}$ per source in FUV ($\sim$ 9.6~ABmag, $\sim$6~e-12 erg~s$^{-1}$~cm$^{-2}$~\AA$^{-1}$). 
%
 GALEX tiles were positioned  avoiding  stars brighter than the above limits in the field
of view; in many regions near the Galactic plane bright stars are numerous, making it impossible
to find a location as large as the GALEX field of view in between too-bright sources,
and causing the known gaps in coverage during the main mission phase, when the FUV detector was
working.

\section{Discussion. The UV sky.}
\label{s_uvsky}

\begin{figure}
\label{f_maps2}
\begin{center}
\includegraphics*[width=6.7cm]{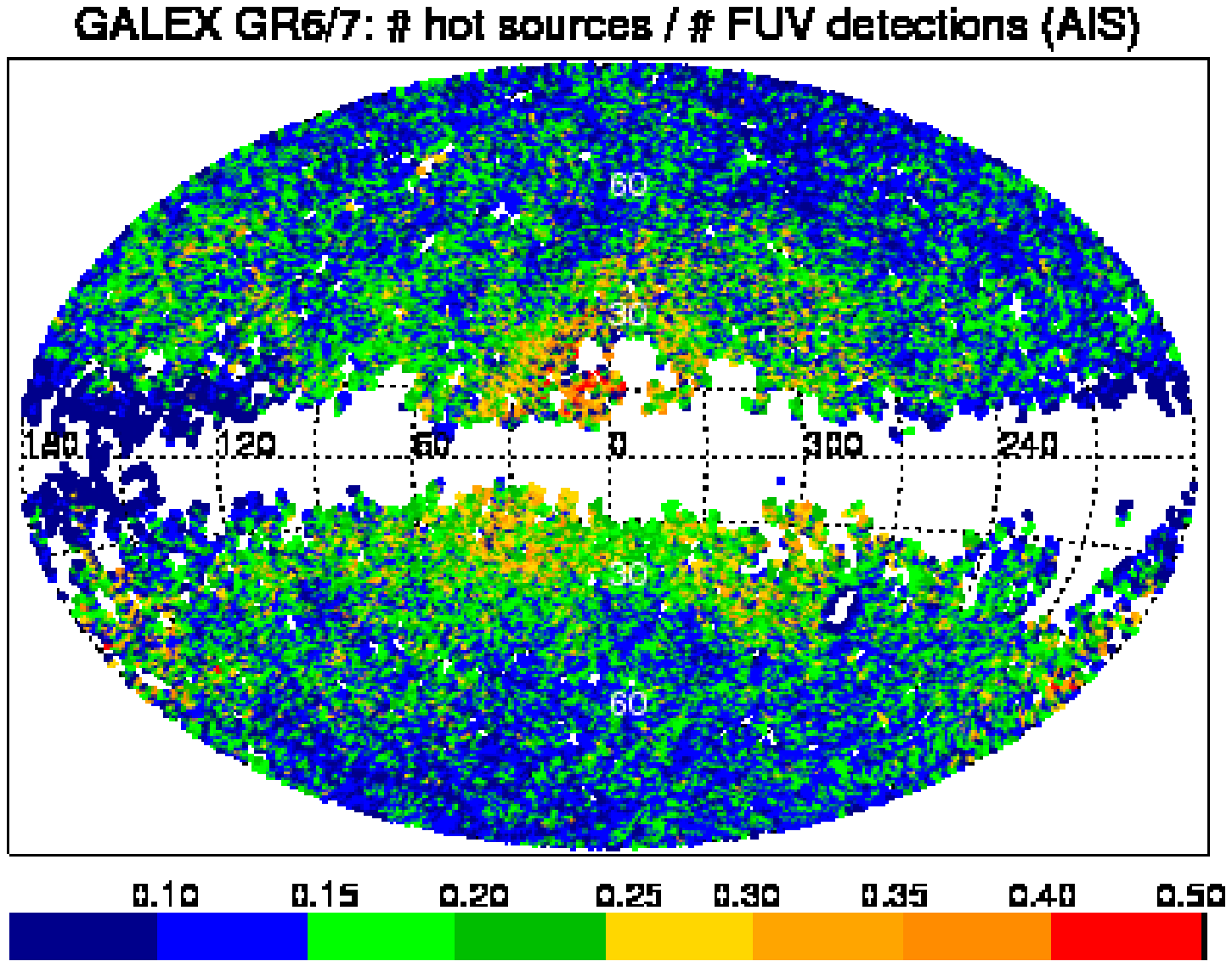} 
\includegraphics*[width=6.7cm]{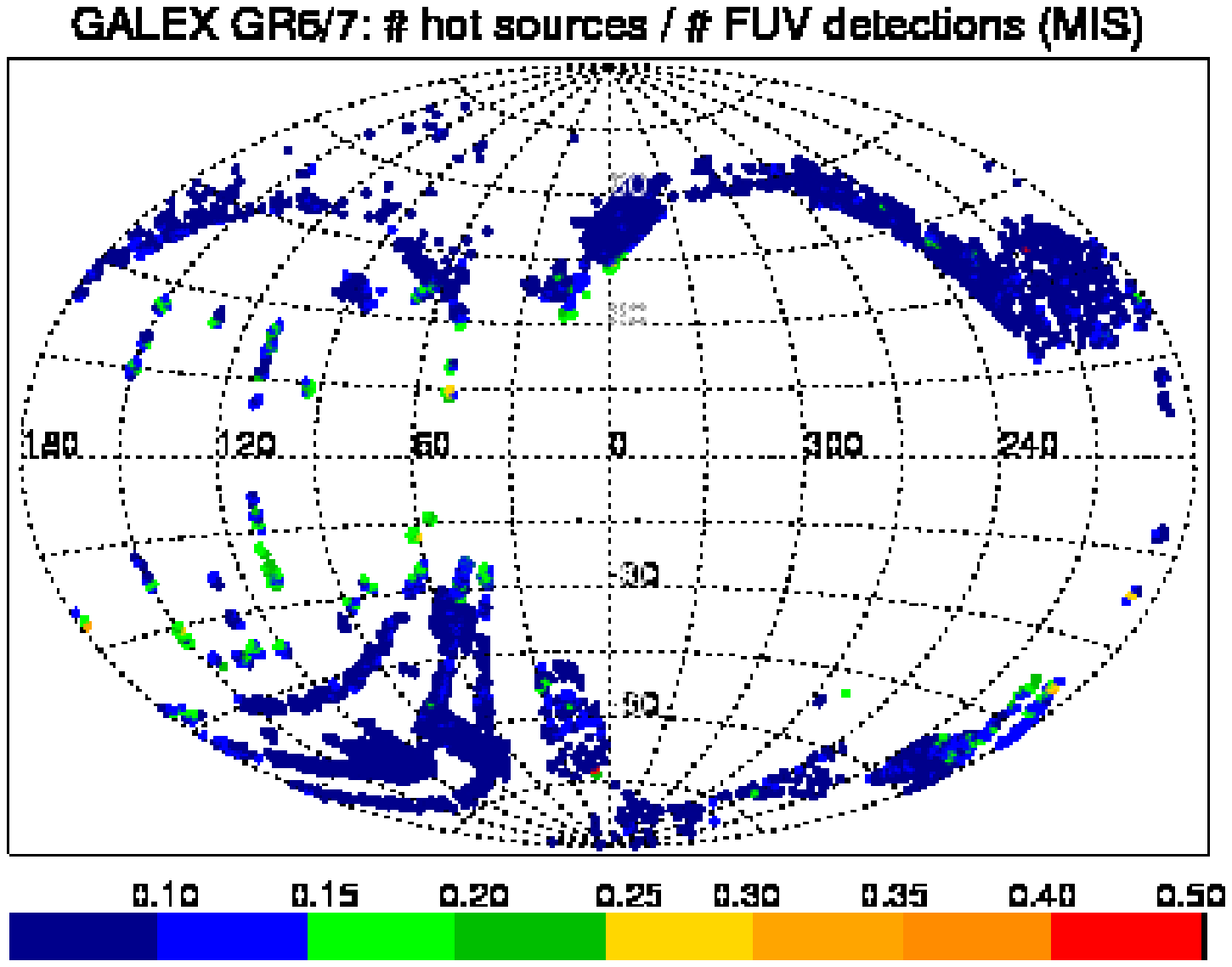} 
\includegraphics*[width=6.7cm]{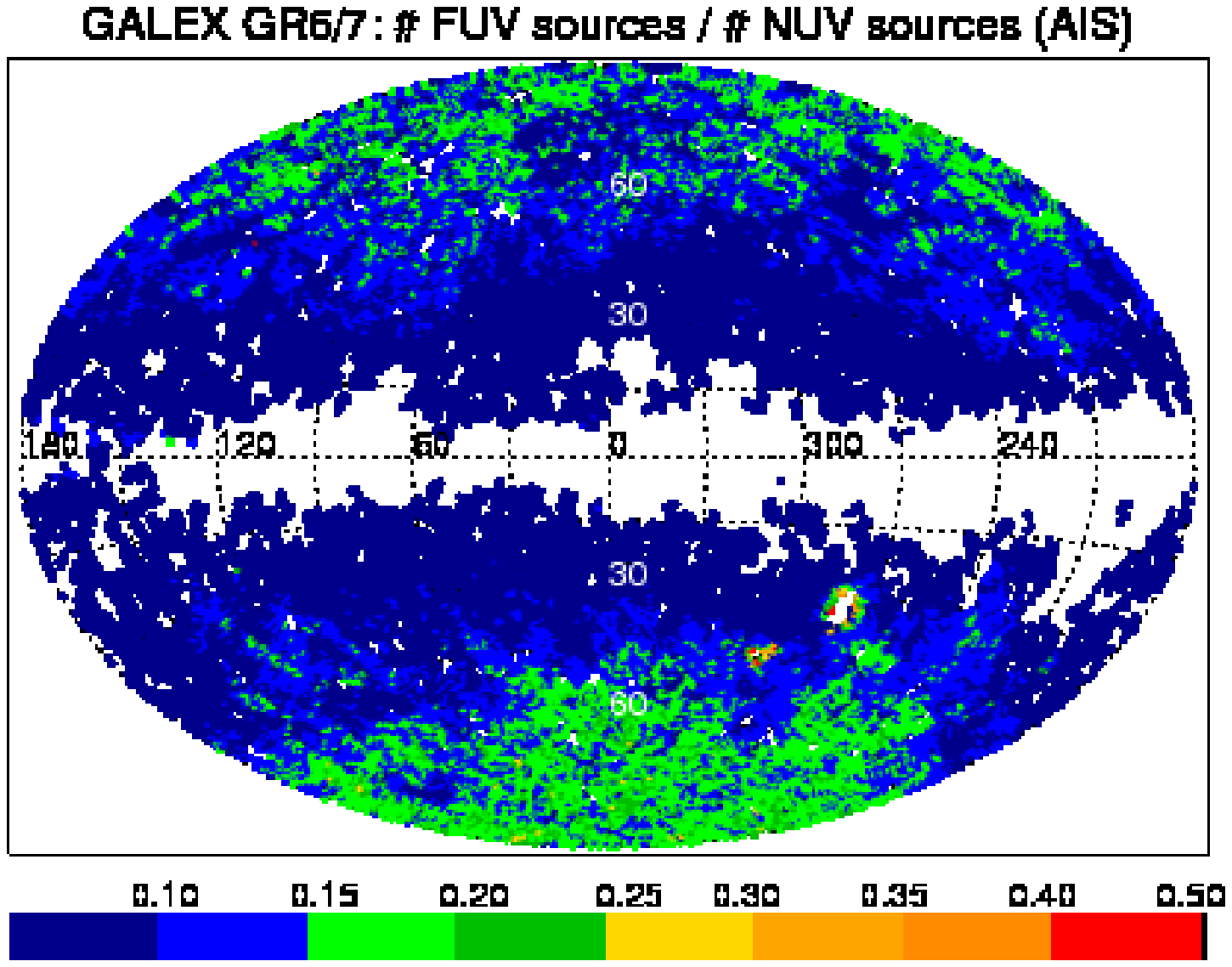} 
\includegraphics*[width=6.7cm]{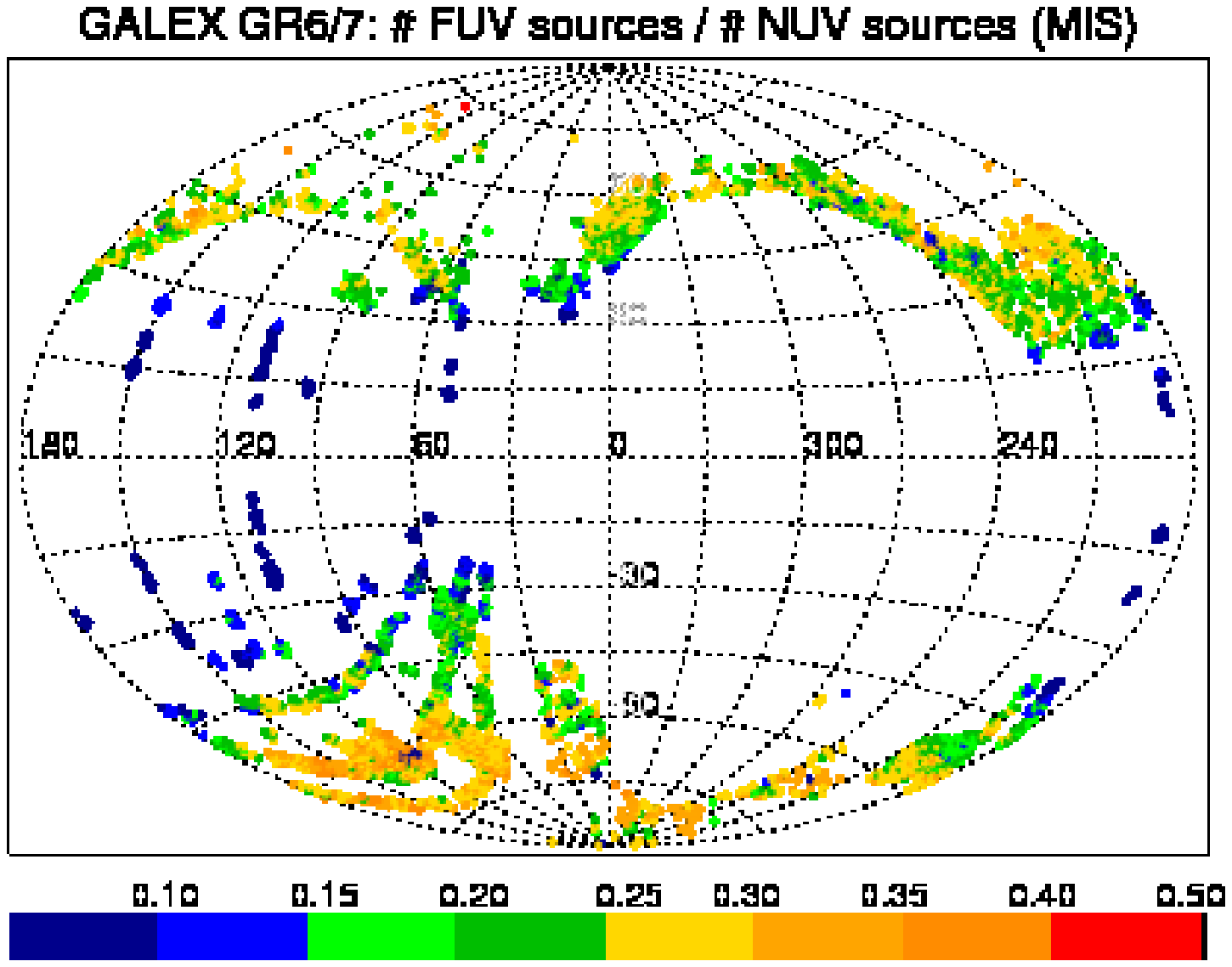} 
\includegraphics*[width=6.7cm]{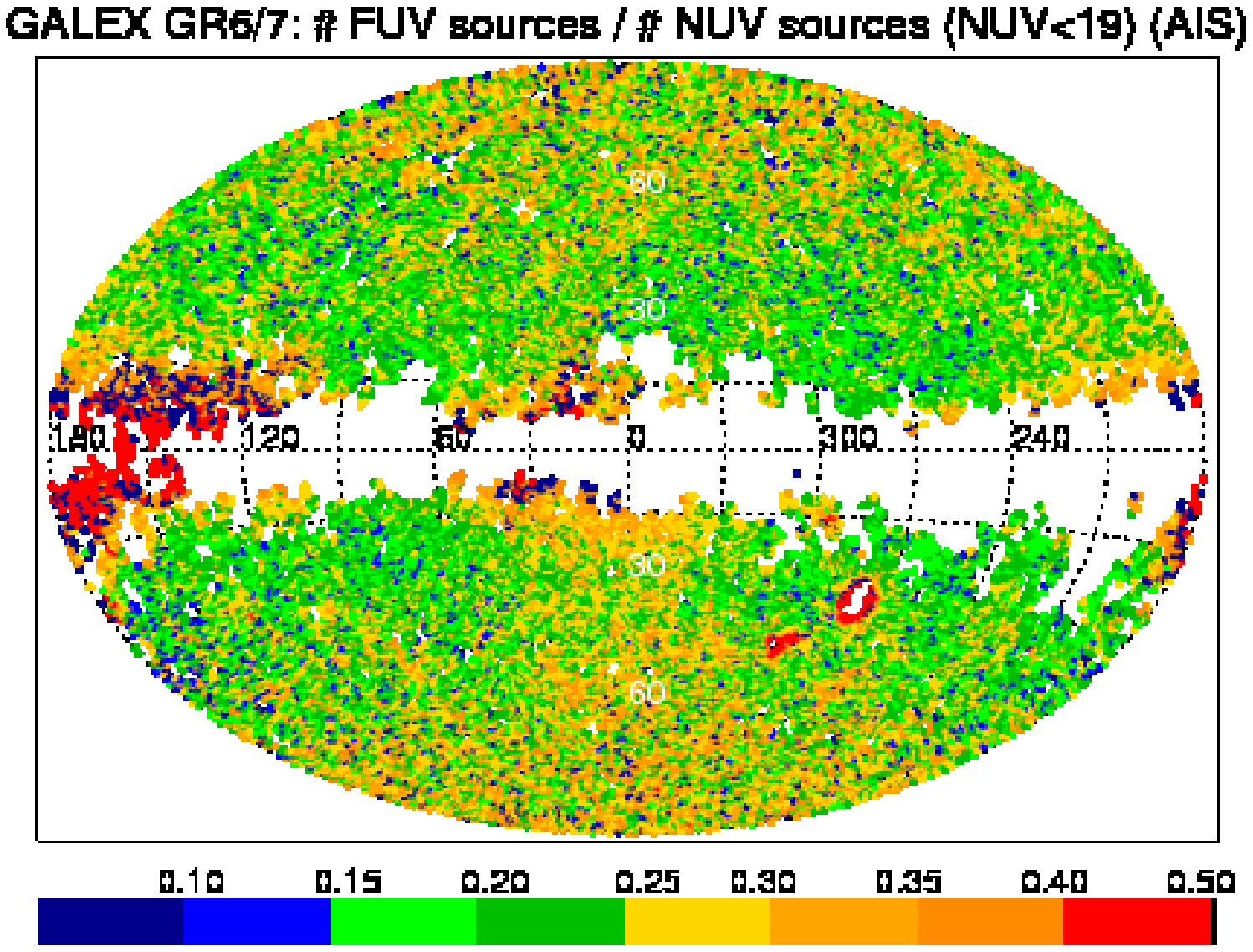} 
\includegraphics*[width=6.7cm]{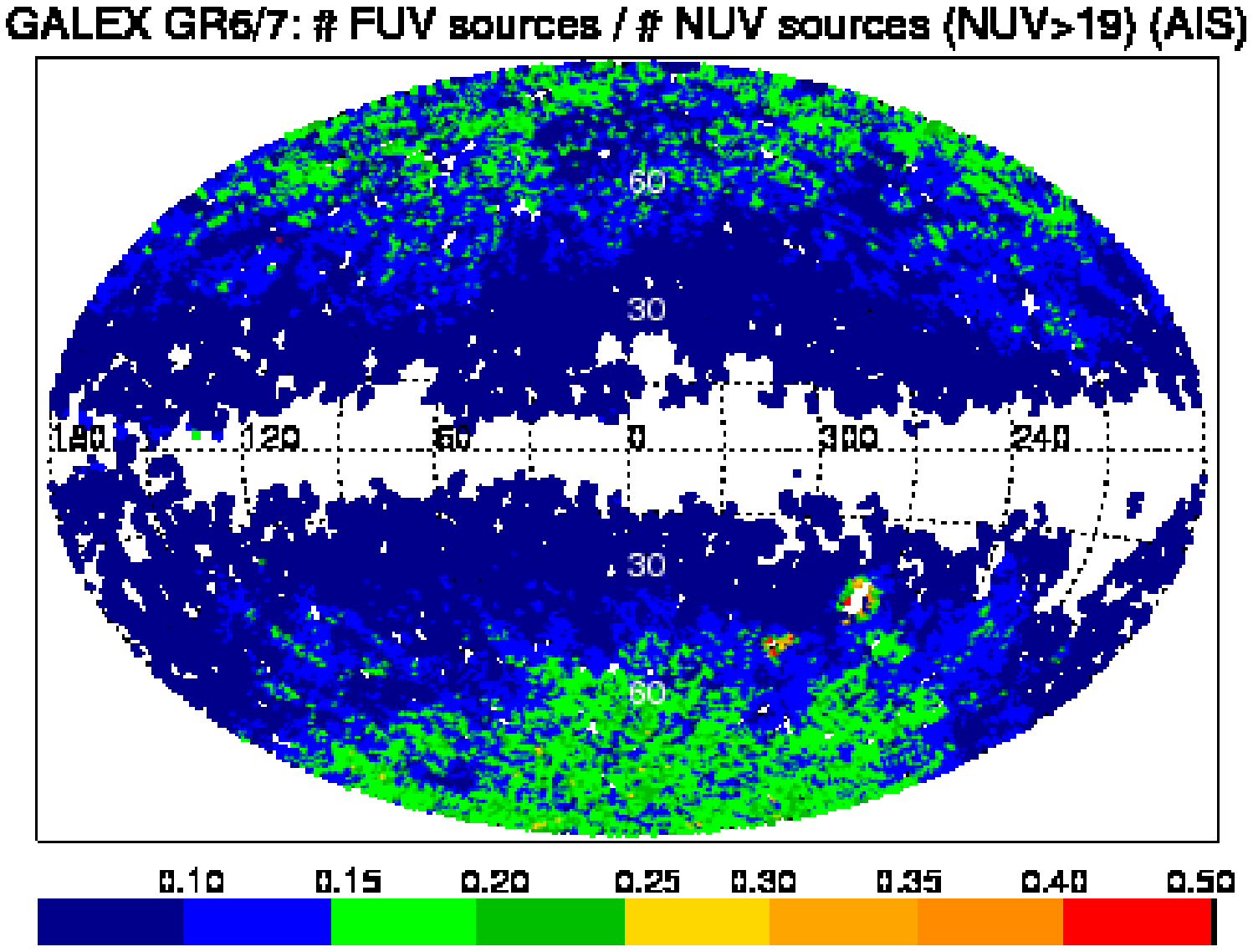} 
\end{center}
\vskip -1.cm 
\caption{
The top maps show  in Galactic coordinates the
fraction of sources with FUV-NUV$\leq$-0.1 (for this map error cuts of 0.3mag are applied to both FUV and NUV mag.s) for AIS and MIS. 
The middle panels show the fraction of NUV sources with significant detection also in FUV, 
over the total number of NUV detections; the bottom panels show
the same fraction, for AIS, separately for sources brighter (left) and fainter (right) than NUV=19mag. 
Bright sources are dominated by Galactic stars, faint sources by extragalactic objects.  
}
\end{figure}

\subsection{Completeness and General Considerations.}
\label{s_caveats}

In any analysis of astrophysical samples we must be concerned with the 
completeness of the catalog, and aware of possible biases. 
 We kept the two catalogs, MIS and AIS, separate because of the $\approx$2~mag difference in 
depth. When density of sources (counts per unit area) is compared among fields,
for example as a function of Galactic latitude and longitude, it is useful to have a sample with
roughly consistent magnitude limits.  
Note, however, that the magnitude limit reached for a typical AIS observation
 (with equal exposure time in both detectors)  differs between FUV and NUV by about 
1~mag (Section 3.1), because  of the instrument sensitivity;
instead, an equal magnitude limit  
in both detectors is reached at MIS depth, for equal exposures, because the sky background is darker in FUV than in NUV 
(see Bianchi 2011 for more details).  Therefore, 
for example, a sample with FUV-NUV=0 will be limited by the FUV-magnitude completeness at AIS depth,   
but a sample with FUV-NUV=-1 would extend in FUV to the same depth as in NUV, one magnitude fainter. 
Instead, at MIS depth, sources with FUV-NUV=0 have the same magnitude limit in FUV and NUV, 
but for ``bluer'' (negative FUV-NUV) sources the completeness limit is set by the NUV magnitude limit,
and viceversa, for redder sources,  by the FUV limit.
It is important to take into account these limits, since different types of sources have 
different magnitude and color distributions, as discussed below. 

\subsection{Density and Distribution of the UV Sources. Dust Extinction}

The catalogs contain sources of various nature: stars, reflecting the density of the 
Milky Way's stellar populations (thin disk, thick disk, halo) with their different 
age and relative content of hot stars, and different geometry (spatial distribution); 
QSOs whose UV colors (and UV$-$optical colors)  depend on
redshift and intensity of emission lines (e.g. Bianchi et al. 2009);
galaxies whose relative UV-to-total flux depends on galaxy type, star-formation rate,
age of the populations, metallicity, inclination (see e.g. Figures 5 and 6 of  Bianchi 2011; Marino et al. 2013, 
2011), as well as on redshift. 

In addition to the intrinsic distributions,  
 UV counts of both Galactic and extragalactic objects are modulated by the foreground extinction from
the Milky Way dust, which is  severe near the Galactic plane (e.g. Figure 2 of Bianchi et al. 2011a).  UV measurements are very sensitive to extinction, since the selective 
extinction \aebv increases steeply at short wavelengths. Therefore, high reddening makes 
observed UV fluxes much fainter than their instrinsic values, effectively limiting the detections
and source counts to brighter intrinsic  magnitudes.  However, while both GALEX FUV and NUV passbands have 
higher \aebv than optical and IR bands, for a Milky-Way type  extinction curve with \rv=3.1 
the extinction coefficients of the two UV bands are very similar, making the FUV-NUV color almost 
reddening-free. This is due to the 2175\AA~ extinction feature being included in the very broad NUV bandpass (see
Bianchi 2011 for more details).  Therefore,  a selection of objects (e.g. hot stars) by 
FUV-NUV color is very robust, and almost unaffected by reddening (as long as the selective extinction is Milky~Way typical).
 Of course, this is not the case for a UV$-$optical color. 
For example, for a Milky~Way-type  extinction curve with \rv$\sim$3.1, \afebv = 8.06 and \anebv = 7.95, hence
E$_{(FUV-NUV)}$/\ebv = 0.11, while for example \eub/\ebv  $\sim$0.70.
However,  in lines of sight where the extinction 
curve is steeper, such as found 
in star-forming sites, E$_{(FUV-NUV)}$/\ebv~ can be 
up to a few magnitudes.  By consequence, assumptions on the type of selective extinction bear severely on the 
results derived from UV magnitudes. 
Coefficients for extinction correction of GALEX magnitudes, A$_{FUV}$/\ebv ~ and A$_{NUV}$/\ebv, for  
different dust assumptions, can be found in Table 2 of Bianchi 2011.

  The relative fraction of Galactic stars over extragalactic objects decreases steeply
towards fainter magnitudes. It is approximately equal at  
AIS depth (Bianchi et al. 2007) although in more detail the Galactic hot-star counts
strongly depend on the Galactic structure, their observed density varying 
with latitude by  a factor of $>$7
 even before  extinction effects are accounted for (Bianchi et al. 2011a).  
 
 The panels in the second  row of Figure \ref{f_maps} and in Figure \ref{f_maps2} 
show the density of sources detected in NUV,
and the fraction of FUV detections over NUV detections, in each field location. 

 The source counts shown in the figures
are estimated within a circle of 0.5$^{\circ}$ radius around each field center.
To obtain density per unit area, they can be re-scaled to an area of one square degree ($\times$ 1.27324). 
  The NUV-source counts (Figure \ref{f_maps}) reflect the effects of foreground extinction and intrinsic source distribution.
The intrinsic distribution of extragalactic sources is independent of sight-line, but their
 number is strongly increasing towards fainter magnitudes.
 Star density increases from the halo to the thick disk and 
to the thin disk, with a marked dependence mainly on Galactic latitude.  Yet we see in the 
source-density maps (Figure \ref{f_maps}) strong variations  with both latitude and longitude. 
This reflects the exinction by Milky Way dust. In fact, there is a remarkable correspondence 
 between this map (density of UV sources) and the Milky Way dust maps, such as those produced by the Planck satellite 
(e.g. Abergel et al. 2011):
the observed density of UV sources is largely anti-correlated with the major dust emission structures, showing the
sensitivity of UV fluxes to extinction.
Some small-scale, patchy inhomogeneities in the counts, e.g. at North Galactic latitudes, are due to longer-than-average 
exposure times as can be seen in the top panels. This is also the cause for counts enhancement
towards the South Galactic pole at longitude $\sim$0. 
When we restrict these plots to observations with more homogeneous exposure times, 
 between 150 and 250~sec for the AIS catalog, and between 1500 and 2500~sec for the MIS catalog, some features 
such as the overdensity at the South  Galactic pole are eliminated.  Comparison between the top-row and 
second-row panels of Figure \ref{f_maps} shows where the overdensities of counts are due to longer exposure times.
Other features are largely due to extinction. 

\subsection{Characteristics of the UV Sources}

The  fraction of FUV detections over NUV detections,
i.e. NUV-detected sources that also have a significant FUV measurement,  shown in the middle 
panels of Figure \ref{f_maps2}, 
 increases towards the Galactic poles, where extinction is 
less and therefore fainter magnitudes are reached in both filters.
In more detail, this trend strongly depends on the magnitude range considered (e.g. Bianchi et al. 2011b).
Hot stars are intrinsically more numerous towards the Milky Way disk, but they are rare. 
Fainter magnitudes include a higher fraction of extragalactic sources. 
This effect is illustrated 
in the bottom panels of Figure \ref{f_maps2}, showing the fraction of 
FUV detections over NUV detections for sources brighter, and fainter, than NUV=19mag. 

The fraction of ``hot'' sources (FUV-NUV$<$-0.1, which corresponds e.g. to stars
with \Teff$\gtrsim$18,000K, the exact limit depending on gravity, metallicity, and model assumptions) is shown in the 
top  panels of Figure \ref{f_maps2}. It is the result of  the combined 
effect of extinction, actual number of hot stars (both increasing towards low latitudes but having
 opposite effects on the counts),
and the relative fraction of extragalactic objects again being different at different magnitudes,
although extragalactic objects mostly have FUV-NUV $>$ 0 (Bianchi 2011).  
As we mentioned previously, the FUV-NUV color is almost reddening-free (for average Milky Way dust type). 
Therefore, reddening has not a big effect on the source selection by color, but it limits 
the source counts, having a large effect on each UV magnitude. By showing the ratio between sources with
negative FUV-NUV and total number of FUV sources, we minimize the influence of 
 count reduction by reddening on the resulting picture (except for the variation of the ratio 
Galactic/extragalactic sources at different limiting magnitudes). 
 
 The difference between the behaviour of Galactic {\it versus} extragalactic samples can be appreciated 
qualitatively by comparing the maps for AIS and MIS (left and right panels respectively in Figures \ref{f_maps} and \ref{f_maps2},
and bottom panels of \ref{f_maps2}), 
the latter being two magnitudes deeper (NUV) and therefore
having a much higher fraction of galaxies and QSOs.  This explains why the relative number of MIS objects with 
negative FUV-NUV is much lower than in the AIS sample, where stars are a substantial fraction. 

 We saw that our catalogs do not have precisely homogeneous exposure times (section \ref{s_surveys}).
When we restrict the plots to data with homogeneous exposures, 
the resulting maps have less uniform coverage (some gaps are introduced), but the overall distribution is very similar to the
maps shown, which include all data; deviations concerning individual fields do not stand  out in the general trends
 when  the whole sky is presented at a glance.

 Finally, to separate the sources into  classes of astrophysical objects, more colors are needed, 
as was shown for example by Bianchi (2009) and Bianchi et al. (2007, 2011a) on earlier, smaller samples
 matched to optical data (see next Section).  
The maps in Figures \ref{f_maps} and  \ref{f_maps2} show for the first time a complete and unbiased 
view of the UV source distribution, without biases and limits of the optical matches.

\section{Matched catalog of UV sources with optical surveys}
\label{s_match}

\begin{figure}
\label{f_histfn}
\begin{center}
\includegraphics*[width=14.cm]{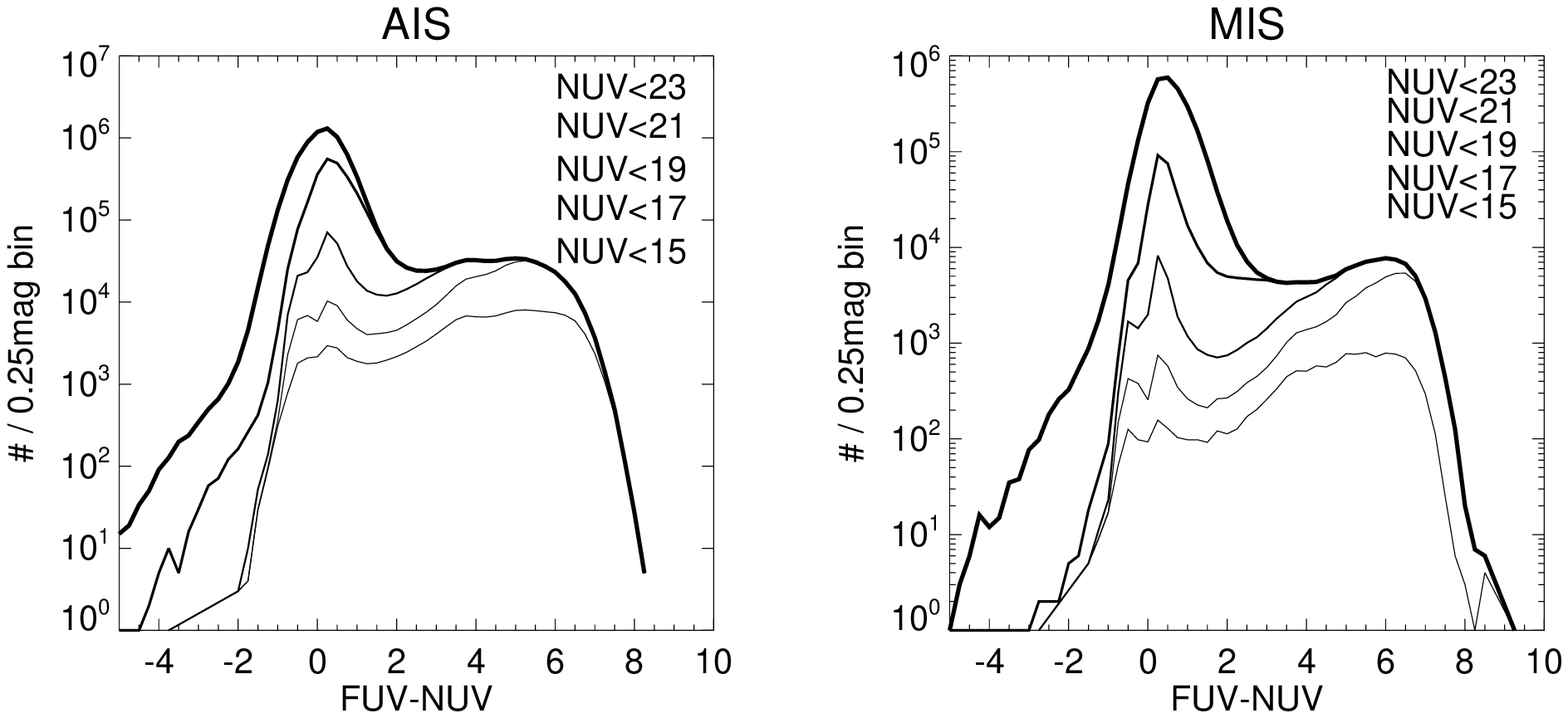}
\includegraphics*[width=14.cm]{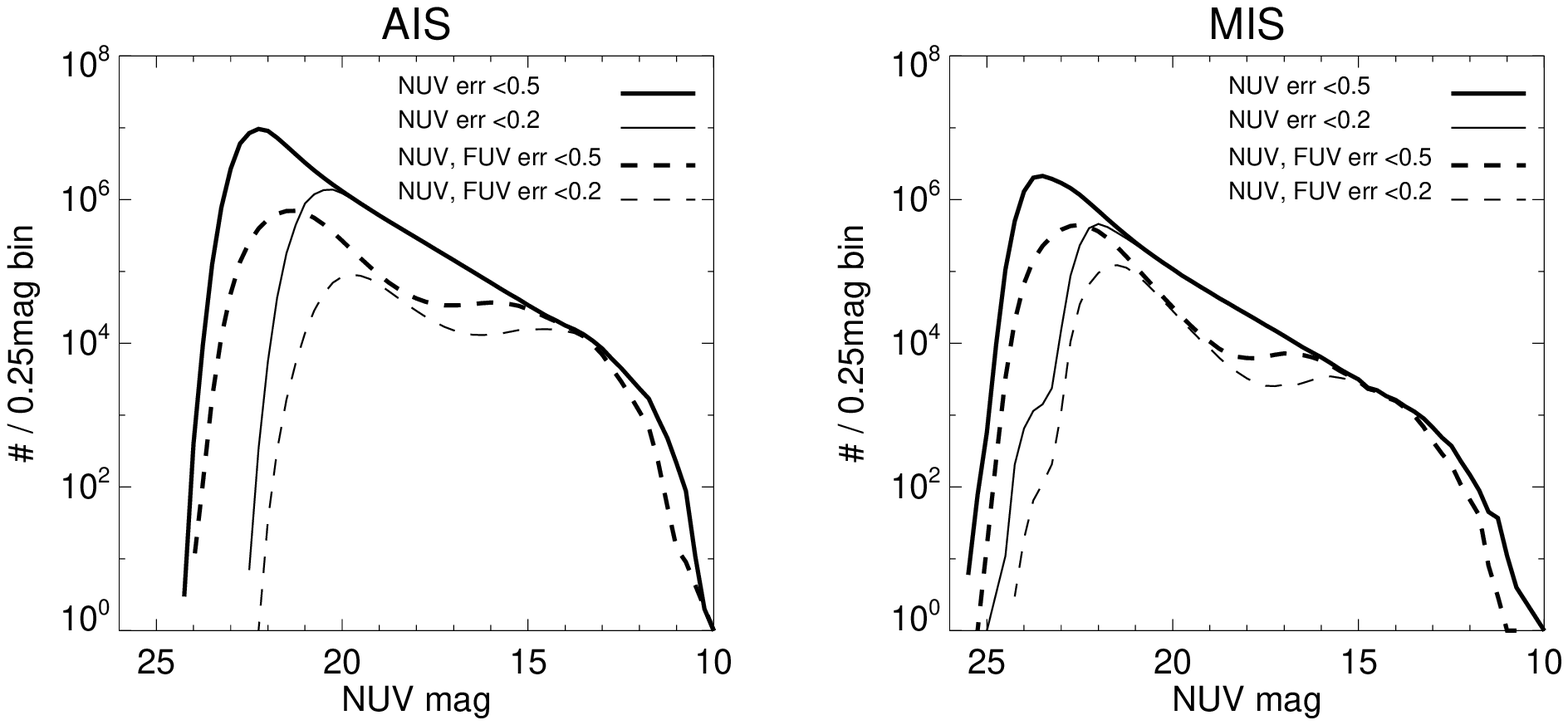}
\end{center}
\caption{Histogram of UV color distribution (top) and NUV mag (bottom) for the MIS and AIS catalogs, with different cuts.
Together with color-color diagrams shown e.g. by Bianchi (2009) and Bianchi et al. (2011a), these diagrams can
be interpreted to characterize at a glance the nature of sources in the UV sky. The bimodal distribution in the 
top histograms is due to Galactic moderately cool stars dominating the redder FUV-NUV peak
 (these are bright and do not suffer much from magnitude cuts),
extragalactic sources (galaxies and QSOs, from FUV-NUV=0 and redder, more abundant at fainter magnitudes), and hot stars (FUV-NUV $<$0, 
very rare). The tail at very negative, unrealistic  FUV-NUV values, is due to sources with large photometric errors, and it is eliminated by 
error cuts (not shown) or magnitude cuts (to some extent equivalent to error cuts).  
The bottom plots show that: (1) a stringent error cut rises the magnitude limit of the sample to brighter magnitudes (thick to thin lines),
as expected, 
and (2) when FUV-detected sources are selected, a bimodal distribution is introduced which is not present in the NUV-detected sample. 
Galactic objects dominate the counts at bright magnitudes, and their relative number decreases at fainter magnitudes.}
\end{figure}

 In order to select samples of objects by astrophysical class with high purity, i.e. minimizing 
contamination from other types of objects, and to perform a detailed analysis and characterization
of their physical parameters, it is useful to match the UV catalogs with databases
at longer wavelengths. 
 In most previous works, GALEX catalogs have been analyzed together with Sloan-Digital-Sky-Survey (SDSS) data,
the widest optical survey easily available since the beginning of the GALEX mission. 
The latter provides 5 optical magnitudes ({\it u g r i z}) which are a useful
 addition to the GALEX FUV and NUV bands, as well as spectra for a subsample of the 
sources. The spectra are useful to validate the  photometrically-derived parameters,
which are available for the 
larger photometric sample, and the object selection, although the spectroscopic sample is serendipitous but not unbiased.  
 Examples are shown e.g. in Bianchi et al. (2011a, 2009, 2007), Treyer et al. (2007), among  others. 

  For science applications concerning Milky Way stellar populations, and the distribution of UV sources 
across the sky, the overlap between the SDSS and GALEX footprints provides insufficient coverage in l,b
to disentangle the effects of  Galactic structure of dust and stellar populations,  
as seen in the bottom-left panel of Figure \ref{f_figure1}.  
For this panel,  the overlap of our GALEX catalogs with SDSS data release 9 (DR9)
has been estimated  by matching the 
locations of our GALEX fields to the SDSS {\it source catalog}: 
those positions where counts of SDSS sources are not zero are deemed to be inside the 
SDSS footprint. This is necessary 
because the SDSS footprint-query tool is currently not working for DR9, and because for previous releases 
we noted that such footprint queries may give both false-positive and false-negative results
(see Bianchi et al. 2011a). 
 The depth of the SDSS data matches well the AIS catalog, while in the GALEX MIS catalog the 
hottest stars, even with very small radii, are detected all the way out to the Galactic halo, in sight-lines of low reddening,  
but these sources are too faint for the SDSS limits (Bianchi et al. 2011a). 

 An important improvement in coverage, and a slight improvement in depth, in terms of optical matches, is 
afforded by the  Pan-STARRS PS1 survey.   Pan-STARRS's PS1 ``3$\pi$ survey'' provides {\it g r i z y} measurements,
 about a half a magnitude deeper than SDSS over a region of the sky
several times larger (Fig. \ref{f_figure1}, bottom-right panel).
 Matched UV-optical catalogs with the 3$\pi$ survey will enable for example a 
conclusive census, and analysis,  of Milky Way hot WDs, extricating effects of stellar evolution 
from geometry of the stellar populations and dust distribution in the Milky Way,
ultimately constraining some still obscure stages of stellar evolution.  We  plan to provide these catalogs 
after the 3$\pi$ survey is completed. 

GSC-II provides some optical magnitudes over the whole sky; 
its nominal depth  matches that of 
the GALEX AIS (the survey with the widest sky coverage), but in practice its completeness is limited to 
 brighter magnitudes and is  not homogeneous (Bianchi et al. 2011b). 
 It is therefore of little use for statistical studies. Yet, it provides to date the only way
 to explore 
the distribution of UV-optical  sources over any direction in the sky, albeit limited to very bright objects, 
possibly supporting and
 complementing studies  of local WD samples (for which proper motions 
and parallax are also available), and providing selection of nearby bright  objects accessible for spectroscopy, 
outside the Pan-STARRS and SDSS coverage.

 We are preparing  matched-source catalogs correlating the final GALEX AIS and MIS catalogs presented here
with SDSS-DR9, GSC-II and  Pan-STARRS 3$\pi$ surveys; these will 
made available from the same web sites as our GALEX catalogs. They will also include corollary data 
such as proper motions and parallaxes when available. 

\section{Summary}
After  eight years of GALEX observations, we have presented here a global view of 
the UV sources across the sky derived from the GALEX  surveys. 
We constructed distilled catalogs of UV sources from the two major surveys, AIS 
containing $\approx$71million sources over $>$22,000 square degrees down to  $\sim$20/21mag (FUV/NUV), and MIS with 
$\approx$ 16.6million sources over $>$2,250 square degrees with a depth of $\sim$22.7mag.  
 The  catalogs presented here  contain ``unique sources'': duplicate measurements
of the same object  
have been eliminated, which greatly facilitates all statistical studies where source counts
of given classes of objects are needed.   The present 
version of the catalogs is limited to observations where both FUV and NUV detectors were exposed, so that any sample
selections by UV color will be unbiased.  
We have shown the significant effect of the Milky Way dust on the UV source counts, 
and global features such as the increase of hot stars from the halo towards the 
Milky Way disk, with a clear dependence on Galactic latitude but also with local structures. 
Distributions of UV magnitudes and FUV-NUV color show that hot stars, cooler stars, and
extragalctic sources (QSOs and galaxies) contribute to different extents in different regimes.

~\\
{\bf Acknowledgements: }
We express our gratitude and appreciation to the GALEX ``Operation and Data Analysis Team'',
who processed and delivered the extremely high data volume yielded by GALEX, 
and to the MAST archive team for making the GALEX data products available in a variety 
of formats, with convenient browsing tools.   
We are  grateful to Ani Thakar who kindly helped us to diagnose issues in the SDSS
footprint queries, which cannot be used.

\clearpage

\appendix

\section{Description of The Catalogs' Columns}

Below we list the tags included in the online catalogs presented in this paper, and 
available at http://dolomiti.pha.jhu.edu (as well as from MAST and SIMBAD in the future).
 The columns of greatest interest in most cases are in bold in the Table below. 
Catalogs with reduced size, including only tags listed in bold-face, are posted on the 
web site for quick download.

\begin{table}[h]
\caption{Catalogs Tags }
\begin{tabular}{ll}
\hline
Tag & Description\\
\hline
photoextractid   &  Pointer to photoExtract Table (identifier of original observation)\\
mpstype	   & which survey  (e.g, "MIS")\\
avaspra	   & R.A. of center of field where object was measured\\
avaspdec   & Decl. of center of field where object was measured\\	
{\bf objid}	   & GALEX identifier for the source\\
{\bf ra	  } & source's  Right Ascension (degrees).\\
{\bf dec }	   & source's Declination  (degrees)\\
{\bf glon }   & source's Galactic longitude  (degrees)\\
{\bf glat }	   & source's Galactic latitude (degrees)\\
tilenum	   & ``tile'' number \\
img	   & image number (exposure \# for \_visits) \\
subvisit   & number of subvisit if exposure was divided\\	
fov\_radius & distance of source from center of the field in which it was measured\\	
type	   & Obs.type (0=single,1=multi)\\
band	   & Band number (1=nuv,2=fuv,3=both)\\
{\bf e\_bv	 }  & E(B-V) Galactic Reddening (from Schlegel et al. 1998 maps)\\
istherespectrum	   & Does this object have a (GALEX) spectrum? Yes (1), No (0) \\
chkobj\_type	   & Astrometry check type\\
{\bf fuv\_mag	}  & FUV calibrated magnitude\\
{\bf fuv\_magerr	 } & FUV calibrated magnitude error \\
{\bf nuv\_mag	 } & NUV calibrated magnitude\\
{\bf nuv\_magerr	}  & FUV calibrated magnitude error \\
fuv\_mag\_auto	   & FUV Kron-like elliptical aperture magnitude\\
fuv\_magerr\_auto	   & FUV RMS error for AUTO magnitude\\
nuv\_mag\_auto	   & NUV Kron-like elliptical aperture magnitude\\
nuv\_magerr\_auto	   & NUV RMS error for AUTO magnitude\\	
\hline
\end{tabular}
\end{table}

\begin{table}[h]
\caption{Catalogs Tags - continued from previous page  }
\begin{tabular}{ll}
\hline
fuv\_mag\_aper\_4	   & FUV Magnitude aperture ( 8 pxl ) \\
fuv\_magerr\_aper\_4  & FUV Magnitude aperture error ( 8 pxl ) \\	
nuv\_mag\_aper\_4	   & NUV Magnitude aperture ( 8 pxl )\\
nuv\_magerr\_aper\_4   &NUV  Magnitude aperture ( 8 pxl ) error\\	
fuv\_mag\_aper\_6	   & FUV  Magnitude aperture ( 17 pxl )\\
fuv\_magerr\_aper\_6  & FUV  Magnitude aperture ( 17 pxl ) error \\	
nuv\_mag\_aper\_6	   & NUV Magnitude aperture ( 17 pxl )\\
nuv\_magerr\_aper\_6  & NUV Magnitude aperture ( 17 pxl ) error\\
{\bf fuv\_artifact }	   & FUV artifact flag (logical OR near source)\\
{\bf nuv\_artifact}	   & NUV artifact flag (logical OR near source)\\
fuv\_flags	   & Extraction flags\\
nuv\_flags	   & Extraction flags\\
fuv\_flux	   & FUV calibrated flux (micro Jansky)\\
fuv\_fluxerr	   & FUV calibrated flux (micro Jansky) error \\
nuv\_flux	   & NUV calibrated flux (micro Jansky)\\
nuv\_fluxerr	   & NUV calibrated flux (micro Jansky) error \\
fuv\_x\_image	   & Object position along x\\
fuv\_y\_image	        & Object position along y\\
nuv\_x\_image   	& Object position along x\\
nuv\_y\_image   	& Object position along y\\
fuv\_fwhm\_image	& FUV FWHM assuming a gaussian core\\
nuv\_fwhm\_image	& NUV FWHM assuming a gaussian core\\
fuv\_fwhm\_world	& FUV FWHM assuming a gaussian core (WORLD units) \\
nuv\_fwhm\_world	& NUV FWHM assuming a gaussian core (WORLD units)\\
nuv\_class\_star	& S/G classifier output \\
fuv\_class\_star	& S/G classifier output \\
{\bf nuv\_ellipticity}	& 1 - B\_IMAGE/A\_IMAGE\\
{\bf fuv\_ellipticity}	& 1 - B\_IMAGE/A\_IMAGE\\
nuv\_theta\_J2000	& Position angle (east of north) (J2000) \\
nuv\_errtheta\_J2000    & Position angle error (east of north) (J2000)\\	
fuv\_theta\_J2000	 & Position angle (east of north) (J2000)\\
fuv\_errtheta\_J2000    & Position angle error (east of north) (J2000)\\	
fuv\_ncat\_fwhm\_image & FUV FWHM\_IMAGE value from -fd-ncat.fits (px)\\	
fuv\_ncat\_flux\_radius\_3	& FUV FLUX\_RADIUS \#3 (-fd-ncat)(px)[0.80]\\
nuv\_kron\_radius& Kron apertures in units of A or B\\	
nuv\_a\_world & Profile RMS along major axis (world units)\\
fuv\_kron\_radius& Kron apertures in units of A or B\\	
fuv\_b\_world & Profile RMS along major axis (world units)\\
\hline
\end{tabular}
\end{table}


\end{document}